\documentclass[twocol]{ametsocV6.2}

\usepackage[T1,OT1]{fontenc}
\DeclareUrlCommand{\DOItext}{%
}

\makeatletter
\newcommand{\setappendixlabel}[2]{%
    \protected@edef\@currentlabel{#1}%
    \label{#2}%
}
\newcommand{\setsubsectionlabel}[1]{%
    \protected@edef\@currentlabel{\thesection\thesubsection}%
    \label{#1}%
}
\makeatother

\title{
Data-driven Reconstruction of the Phase Dynamics for the El Ni\~{n}o--Southern Oscillation Reveals Seasonality and Statistical Synchronization with the Annual Cycle
}

\authors{
Takahiro Arai \correspondingauthor{Takahiro Arai, araitak@jamstec.go.jp}
and Yoji Kawamura
}

\affiliation{{Center for Mathematical Science and Advanced Technology, Japan Agency for Marine-Earth Science and Technology, Yokohama 236-0001, Japan}}

\abstract{
In this study, we treat the El Ni\~{n}o-Southern Oscillation (ENSO) as a rhythmic system periodically forced by the annual cycle.
We examine ENSO event seasonality and synchronization between the annual and ENSO cycles.
We compute phase time series for both cycles from sea surface temperature data using a data-driven Koopman operator approach.
We then use Bayesian inference to reconstruct phase equations describing how the phases of the two oscillators evolve.
The reconstructed two-variable phase equations show that the annual cycle modulates the instantaneous frequency of the fundamental ENSO mode.
Stochastic simulations of these equations incorporating the estimated stationary noise indicate that El Ni\~{n}o and La Ni\~{n}a events tend to last longer when they start in boreal summer and that the variability of their durations is greater when they start in boreal spring.
Furthermore, the averaged form of the phase equations indicates statistical $3:1$ phase locking between the annual and ENSO cycles, whereas no statistical phase locking is found for the other $m:n$ ratios.
}

\begin{document}

\maketitle

\section{Introduction}
\label{sec:introduction}

\par
El Ni\~{n}o-Southern Oscillation (ENSO) is a prominent mode of interannual variability arising from ocean-atmosphere coupling in the tropical Pacific and is a major driver of global interannual climate variability \citep{bjerknes_atmospheric_teleconnections_1969,rasmusson_variations_tropical_sea_surface_temperature_1982,neelin_enso_theory_1998}.
ENSO-related changes in sea surface temperature (SST), trade winds, and convective activity influence atmospheric circulation and precipitation far beyond the tropical Pacific \citep{horel_planetary_scale_atmospheric_phenomena_1981,ropelewski_global_regional_precipitation_patterns_1987}.
Therefore, understanding ENSO dynamics is important for diagnosing climate variability, assessing predictability, and evaluating climate models \citep{timmermann_el_nino_southern_oscillation_complexity_2018,vialard_enso_recharge_oscillator_conceptual_model_2025}.

\par
The mechanisms underlying the development, maturation, and termination of ENSO events have been studied using theoretical models such as the recharge oscillator model \citep{jin_equatorial_ocean_recharge_paradigm_1997,burgers_simplest_enso_recharge_oscillator_2005,vialard_enso_recharge_oscillator_conceptual_model_2025}, linear stochastic models \citep{penland_optimal_growth_1995,thompson_linear_stochastic_dynamical_model_enso_2001,kido_understanding_interplay_enso_2023,lien_cyclostationary_linear_inverse_models_2025,yasuda_stochastic_thermodynamic_constraint_2026}, and intermediate coupled ocean--atmosphere models \citep{zebiak_model_el_nino_southern_oscillation_1987,battisti_dynamics_thermodynamics_warming_event_1988,wang_review_enso_theories_2018}.
The ENSO period varies over approximately 2--7~years, whereas El Ni\~{n}o events tend to develop during boreal summer and autumn and peak in boreal winter.
This seasonality has been discussed in terms of annual--ENSO synchronization \citep{jin_el_nino_devils_staircase_1994,minobe_generation_interannual_interdecadal_climate_oscillations_2004,stein_phase_synchronization_2011,stein_enso_seasonal_synchronization_theory_2014}.
Previous studies reported statistical $2:1$ phase locking between the annual and ENSO cycles \citep{stein_phase_synchronization_2011,stein_enso_seasonal_synchronization_theory_2014}.
\footnote{
    \citet{stein_phase_synchronization_2011,stein_enso_seasonal_synchronization_theory_2014} described a statistical preference for certain phase differences without frequency locking as partial phase synchronization.
    Here, following \citet{schafer_synchronization_1999}, we call this behavior statistical phase locking.
    }
Seasonal modulation of the ENSO growth rate has been proposed as a mechanism that may explain ENSO seasonality \citep{stein_seasonal_synchronization_enso_events_2010,an_linear_solutions_frequency_amplitude_modulation_2011,stein_enso_seasonal_synchronization_theory_2014,chen_dynamics_enso_phase_locking_2022}.
Despite these studies, observations have yet to clarify how the annual cycle modulates the ENSO cycle or how this modulation affects ENSO seasonality and annual--ENSO synchronization.

\par
We model the annual and ENSO cycles as coupled rhythmic systems and represent each rhythm by a single phase variable.
We reconstruct their phase equations from SST observations to quantify annual modulation of ENSO.
Using these phase equations, we evaluate ENSO event seasonality and synchronization with the annual cycle.
This analysis relies on a method for extracting phase variables that represent the annual and ENSO cycles from observational data and a framework for reconstructing their phase dynamics from the resulting phase time series.
We use a data-driven approach based on Koopman operator theory to extract the phases and on phase reduction theory to reconstruct the phase dynamics.

\par
Koopman operator theory provides a framework for describing the time evolution of observables in nonlinear dynamical systems using linear operators \citep{koopman_hamiltonian_1931,mezic_spectral_2005}.
Methods such as dynamic mode decomposition (DMD) and extended DMD construct finite-dimensional Koopman operator approximations and estimate their spectral properties from observations \citep{rowley_spectral_nonlinear_flows_2009,schmid_dynamic_mode_decomposition_2010,williams_data_driven_2015,giannakis_data_driven_2019,brunton_modern_2022,brunton_data_driven_2022,ichinaga_pydmd_2024}.
\citet{froyland_spectral_2021} identified annual and ENSO oscillatory modes by computing approximate Koopman eigenfunctions from SST data using a framework also applied to the Madden--Julian oscillation (MJO) \citep{lintner_identification_2023} and the quasi-biennial oscillation (QBO) \citep{valva_qbo_2025}.
Their method combines delay-coordinate maps and kernel methods to approximate the Koopman generator, separating the fundamental ENSO mode from the annual cycle and combination modes \citep{giannakis_nonlinear_2012,giannakis_data_driven_2019,das_reproducing_2021,froyland_spectral_2021}.
The fundamental mode can therefore be regarded as a pure ENSO mode.
The ENSO eigenfunction yields a rectified phase time series analogous to that of a harmonic oscillator.

\par
These phase time series allow us to quantify how coupling modulates instantaneous frequencies within a dynamical systems framework.
Phase reduction theory provides a framework in which each rhythm is represented by a single phase variable \citep{winfree_geometry_1980,kuramoto_chemical_1984,acebron_kuramoto_2005,nakao_phase_2016,pietras_network_2019}.
The phase dynamics are described by coupled phase equations.
For example, in the absence of coupling, the phase of each rhythm advances at a constant rate; coupling modulates the oscillators’ instantaneous frequencies.
Data-driven approaches have been developed to reconstruct these phase equations from time-series data \citep{rosenblum_detecting_2001,miyazaki_determination_2006,tokuda_inferring_2007,kralemann_phase_2008,stankovski_inference_2012,duggento_dynamical_2012,ota_direct_2014,cestnik_inferring_2018,yawata_phase_2024,arai_extracting_2022,arai_setting_2025,arai_data_driven_2026,matsuki_isochron_free_2026}.
Applications include quantifying selective attention in calling frogs \citep{ota_interaction_2020} and respiratory modulation of heart rate \citep{stankovski_inference_2012,kralemann_vivo_2013}, and analyzing coordination and rhythm control during human walking \citep{funato_evaluation_2016,arai_interlimb_2024,furukawa_bayesian_2025}.
We therefore use this framework to reconstruct phase equations from the phase time series and quantify annual forcing on ENSO.

\par 
Combining a data-driven Koopman operator approach with Bayesian inference of phase equations allows us to quantify ENSO phase dynamics under annual forcing.
The overall procedure is as follows.
First, we compute Koopman eigenfunctions for the annual cycle and the fundamental 4-year ENSO mode from SST data and then compute their phase time series.
We then use these time series to reconstruct equations for two coupled phase oscillators.
The two phase oscillators represent the annual and ENSO cycles, respectively.
We use two forms of the phase equations in our analysis: a form expressed in terms of the two phase variables and an averaged form expressed in terms of the $m:n$ phase difference.
The two-variable form can be reconstructed from the phase time series. 
The averaged form can be obtained either by averaging the two-variable form or by reconstructing it directly from the same time series.
We use the two-variable form to evaluate the seasonal dependence of the durations of ENSO events and the averaged form to evaluate the $m:n$ synchronization properties.

\par
This paper is organized as follows.
In Sec.~\ref{sec:method}, we describe the reconstruction of the phase dynamics from SST data.
In Sec.~\ref{sec:results}, we present the results of our data-driven analyses of the phase dynamics.
In Sec.~\ref{sec:discussion}, we summarize and discuss our results.
Methodological details and additional results appear in Appendices~\ref{appendix_A}--\ref{appendix_G}.

\section{Methods}
\label{sec:method}

\par
This section describes the methods used to quantify the phase dynamics.
In Sec.~\ref{subsec:operator_theoretic_formalism}, we compute annual and ENSO phase time series from SST fields using a data-driven Koopman operator approach~\citep{froyland_spectral_2021}.
In Sec.~\ref{subsec:phase_dynamics_reconstruction}, we reconstruct phase dynamics from these time series.
In Sec.~\ref{subsec:averaging_method}, we introduce the averaging method for deriving the averaged form of the phase equations describing $m:n$ synchronization.
In Sec.~\ref{subsec:direct_averaged_phase_inference}, we present an alternative method that reconstructs this averaged form directly from the phase time series.

\subsection{Operator-theoretic framework for computing the phases of the annual and ENSO cycles}
\setsubsectionlabel{subsec:operator_theoretic_formalism}

\par
We consider a dynamical system with state $\omega(t) \in \Omega$, where $\Omega$ denotes the state space.
The state evolves according to $\omega(t)=\Phi^t(\omega_0)$, where $\Phi^t:\Omega\to\Omega$ denotes the time-$t$ flow map.
An observable $f:\Omega\to\mathbb{C}$ maps the state $\omega(t)$ to the observed value $f(\omega(t))$.
The time evolution of an observable $f$ can be described either by evolving the state in $\Omega$ and then evaluating $f$, or by composing $f$ with $\Phi^t$ and evaluating the resulting function at the initial state:
\begin{align}
    \label{eq:observable_f}
    f(\omega(t_0+t))
    = f(\Phi^t(\omega(t_0)))
    = (f\circ\Phi^t)(\omega(t_0)).
\end{align}

\par
In Koopman operator theory, the Koopman operator $U^t$, which describes the evolution of observables over time $t$, is defined by~\citep{koopman_hamiltonian_1931,brunton_modern_2022}
\begin{align}
    \label{eq:Koopman_operator_U}
    (U^t f)(\omega(t_0))
    = (f\circ\Phi^t)(\omega(t_0))
    = f(\omega(t_0+t)).
\end{align}
Thus, the Koopman operator describes the dynamics through its action on observables rather than through the evolution of states in $\Omega$.
Let $g_j:\Omega\to\mathbb{C}$ be a Koopman eigenfunction satisfying
\begin{align}
    \label{eq:Koopman_eigenfunction}
    U^t g_j= \mathrm{e}^{t\nu_j} g_j,
\end{align}
where $\rho_j(t)=\mathrm{e}^{t\nu_j}$ is the eigenvalue of $U^t$.
The real and imaginary parts of $\nu_j$ represent the growth or decay rate and the eigenfrequency, respectively.
Assuming that the observable $f$ can be expanded in Koopman eigenfunctions, we obtain
\begin{align}
    \label{eq:Koopman_observable_expansion}
    f(\omega)=\sum_j c_j g_j(\omega),
\end{align}
where $c_j$ are the Koopman modes associated with the scalar-valued observable $f$.
For details of the computation, see \citet{froyland_spectral_2021}.

\par
For continuous-time dynamics, infinitesimal evolution of observables is described by the Koopman generator $V$, defined by
\begin{align}
    \label{eq:Koopman_generator}
    Vf=\lim_{t\to0}\frac{U^t f-f}{t},
\end{align}
for observables $f$ for which this limit exists \citep{koopman_hamiltonian_1931,brunton_modern_2022}.
The Koopman operator and its generator are related by
\begin{align}
    U^t f=\mathrm{e}^{tV}f, 
\end{align}
where $\mathrm{e}^A = \sum_k \frac{1}{k !} A^k$.
The eigenfunction $g_j$ evolves according to
\begin{align}
    e^{tV}g_j=e^{t\nu_j}g_j.
\end{align}
Spectral analysis of the Koopman generator (operator) identifies observables $\{g_j\}_j$ associated with oscillatory behavior in the dynamics.
If a few Koopman eigenfunctions approximate the observable well, they compactly represent the temporal behavior of interest.

\par 
We compute the eigenfunctions using the data-driven Koopman operator approach.
This approach approximates the Koopman generator and its eigenfunctions from SST fields without explicit governing equations for the underlying dynamics.
The computational procedure is summarized in Appendix~\ref{appendix_A}.

\par
Koopman generator eigenfunctions associated with the annual cycle, the fundamental ENSO mode with a period of approximately 4 years, a 3-year ENSO mode, and combination modes of the annual and ENSO cycles have been identified~\citep{froyland_spectral_2021}.
To investigate the phase interaction between the annual and fundamental ENSO cycles, we focus on the corresponding eigenfunctions and define their phases from their complex arguments.
The corresponding eigenfunctions can be identified by their eigenfrequencies,
\begin{align}
    \label{eq:koopman_eigenfrequency_annual}
    \frac{\operatorname{Im}\nu_{\mathrm{a}}}{2\pi}
    &\approx 1~[\mathrm{cycle / year}], \\
    \label{eq:koopman_eigenfrequency_enso}
    \frac{\operatorname{Im}\nu_{\mathrm{e}}}{2\pi}
    &\approx 0.25~[\mathrm{cycle / year}],
\end{align}
for the annual and ENSO cycles, respectively.
We denote the eigenpairs associated with the annual and fundamental ENSO cycles by $(g_{\mathrm{a}},\nu_{\mathrm{a}})$ and $(g_{\mathrm{e}},\nu_{\mathrm{e}})$, respectively.
Hereafter, the ENSO cycle refers to the cycle represented by the fundamental ENSO mode.

\par
The complex arguments of the corresponding Koopman eigenfunctions serve as phase variables for the annual and ENSO cycles.
We introduce the sampling times $t_\ell=\ell\Delta t$ and denote the state at time $t_\ell$ by $\omega_\ell=\omega(t_\ell)$, where $\ell=0,1,\ldots,N-1$ and $\Delta t$ is the sampling interval.
The eigenfunctions $g_{\mathrm{a}}(\omega_\ell)$ and $g_{\mathrm{e}}(\omega_\ell)$ can be expressed in polar form in terms of their amplitudes $R_{\mathrm{a}}$ and $R_{\mathrm{e}}$ and phases $\theta_{\mathrm{a}}$ and $\theta_{\mathrm{e}}$, respectively.
\begin{align}
    \label{eq:eigenfunction_a_polar}
    g_{\mathrm{a}}(\omega_\ell) &= R_{\mathrm{a}}(t_\ell) \exp[\mathrm{i} \theta_{\mathrm{a}}(t_\ell)], \\
    \label{eq:eigenfunction_e_polar}
    g_{\mathrm{e}}(\omega_\ell) &= R_{\mathrm{e}}(t_\ell) \exp[\mathrm{i} \theta_{\mathrm{e}}(t_\ell)].
\end{align}
We define the phases of the annual and ENSO cycles as the complex arguments of the corresponding eigenfunctions:
\begin{align}
    \label{eq:phase_a_from_eigenfunction}
    \theta_{\mathrm{a}}(t_\ell) &= \arg \left[ g_{\mathrm{a}} (\omega_\ell)\right], \\
    \label{eq:phase_e_from_eigenfunction}
    \theta_{\mathrm{e}}(t_\ell) &= \arg \left[ g_{\mathrm{e}} (\omega_\ell)\right].
\end{align}
These phases allow us to describe the annual and ENSO cycles as rhythmic systems.

\par
For Bayesian inference \citep{bishop_pattern_2006} based on the models specified in Secs.~\ref{subsec:phase_dynamics_reconstruction} and \ref{subsec:direct_averaged_phase_inference}, we construct datasets using midpoint approximations of the phases.
We also apply an amplitude filter based on the amplitude of the ENSO eigenfunction
\begin{align}
    R_{\mathrm e}(t_\ell):=|g_{\mathrm e}(\omega_\ell)|,
\end{align}
using the threshold $\tilde{R}_{\mathrm e}$ to retain phase data satisfying
\begin{align}
    \label{eq:active_condition}
    R_{\mathrm e}(t_\ell) \geq \tilde{R}_{\mathrm e}.
\end{align}
Otherwise, we regard ENSO states as inactive.
In this low-amplitude regime, eigenfunction values lie near the phase singularity at $g_{\mathrm e}=0$, making the phase sensitive to small perturbations \citep{guckenheimer_isochrons_1975,mauroy_extreme_2015}.
Accordingly, if the condition in Eq.~(\ref{eq:active_condition}) is not satisfied, we treat the phase as unreliable and exclude it from the datasets used for Bayesian inference.
The procedures for constructing these datasets are described in Appendices~\ref{appendix_B} and \ref{appendix_C}.

\par
We construct phase composites to visualize representative SST fields at different stages of the annual and ENSO cycles.
We consider eight stages for each cycle ($S=8$).
The phase composites for the annual and ENSO cycles are calculated from monthly mean SST fields and monthly SST anomaly (SSTA) fields, respectively.
The bin $\Theta_j$ of phase angles is defined by
\begin{align}
    \label{eq:phase_bin}
    \Theta_j
    =
    \left[
    \frac{2\pi(j-\frac32)}{S},
    \frac{2\pi(j-\frac12)}{S}
    \right),
    \qquad
    (j=1,\ldots,S),
\end{align}
where $\Theta_j$ is centered at \(2\pi(j-1)/S\), with phase angles taken modulo $2\pi$.

\par 
The sets of time indices assigned to stage $j$ of the annual and ENSO cycles are defined by
\begin{align}
\label{eq:phase_composite_indices_annual}
D_j^{\mathrm{a}}
&=
\left\{
 \ell
 \;\middle|\;
 \theta_{\mathrm{a}}(t_\ell)\in \Theta_j,
 \; R_{\mathrm{e}}(t_\ell) \geq \tilde{R}_{\mathrm{e}}
\right\},
\\
\label{eq:phase_composite_indices_enso}
D_j^{\mathrm{e}}
&=
\left\{
 \ell
 \;\middle|\;
 \theta_{\mathrm{e}}(t_\ell)\in \Theta_j,
 \; R_{\mathrm{e}}(t_\ell) \geq \tilde{R}_{\mathrm{e}}
\right\}.
\end{align}
States not satisfying Eq.~(\ref{eq:active_condition}) are classified as inactive, and their time indices $\ell$ are excluded from the index sets $D_j^{\mathrm{a}}$ and $D_j^{\mathrm{e}}$.

\par
Following \citet{froyland_spectral_2021}, we visualize representative SST and SSTA fields at each stage $j$ of the annual and ENSO cycles using the phase composites \citep{alexander_kernel_analog_forecasting_2017} defined as follows:
\begin{align}
\label{eq:phase_composite_annual}
\bar{\bm{x}}_j^{\mathrm{a}}
&=
\frac{1}{|D_j^{\mathrm{a}}|}
\sum_{\ell\in D_j^{\mathrm{a}}}
\bm{x}(t_\ell),
\\
\label{eq:phase_composite_enso}
\bar{\bm{x}}_j^{\mathrm{e}}
&=
\frac{1}{|D_j^{\mathrm{e}}|}
\sum_{\ell\in D_j^{\mathrm{e}}}
\bm{x}_{\mathrm{SSTA}}(t_\ell),
\end{align}
where $|D_j^{\mathrm{a}}|$ and $|D_j^{\mathrm{e}}|$ denote the numbers of elements in the respective index sets.
\footnote{
    The selection procedure of \citet{froyland_spectral_2021} keeps $|D_j^{\mathrm a}|$ and $|D_j^{\mathrm e}|$ constant across $j$.
    Here, we retain samples that satisfy the condition in Eq.~(\ref{eq:active_condition}).
    Consequently, the sample counts can vary with $j$.
}
Here, $\bm{x}(t_\ell)$ and $\bm{x}_{\mathrm{SSTA}}(t_\ell)$ denote snapshots at time $t_\ell$ of the monthly mean SST field and the monthly SSTA field, respectively.
The phase composites $\bar{\bm{x}}_j^{\mathrm a}$ and $\bar{\bm{x}}_j^{\mathrm e}$ provide representative SST and SSTA fields for stage $j$, respectively.

\subsection{Data-driven reconstruction of the phase dynamics}
\setsubsectionlabel{subsec:phase_dynamics_reconstruction}

\par
We quantify the phase dynamics of the annual and ENSO cycles by reconstructing stochastic phase equations for two coupled phase oscillators from the phase time series \citep{stankovski_inference_2012,duggento_dynamical_2012}.
We consider the following two-variable form of these equations:
\begin{align}
\label{eq:phase_equation_annual}
\dot{\theta}_{\mathrm{a}}(t)
&= Q_{\mathrm{a}}(\theta_{\mathrm{a}},\theta_{\mathrm{e}}) + \eta_{\mathrm{a}}(t),
\\
\label{eq:phase_equation_enso}
\dot{\theta}_{\mathrm{e}}(t)
&= Q_{\mathrm{e}}(\theta_{\mathrm{a}},\theta_{\mathrm{e}}) + \eta_{\mathrm{e}}(t),
\end{align}
where $\boldsymbol{\eta}(t):=(\eta_{\mathrm{a}}(t),\eta_{\mathrm{e}}(t))^{\mathrm T}$ is a noise vector satisfying
\begin{align}
\label{eq:phase_noise_mean}
\left\langle
    \boldsymbol{\eta}(t)
\right\rangle
&=
\boldsymbol{0},
\\
\label{eq:phase_noise_covariance}
\left\langle
    \boldsymbol{\eta}(t)
    \boldsymbol{\eta}(t')^{\mathrm T}
\right\rangle
&=
{\rm E}\,\delta(t-t').
\end{align}
Here, $\delta(t-t')$ denotes the Dirac delta function.
The functions $Q_{\mathrm{a}}(\theta_{\mathrm{a}},\theta_{\mathrm{e}})$ and $Q_{\mathrm{e}}(\theta_{\mathrm{a}},\theta_{\mathrm{e}})$, which are scalar-valued and $2\pi$-periodic in each phase variable, describe the deterministic part of the phase dynamics  \citep{winfree_geometry_1980,kuramoto_chemical_1984,nakao_phase_2016,pietras_network_2019}.

\par 
To formulate the regression problem for the phase equations, we parameterize $Q_{\mathrm{a}}(\theta_{\mathrm{a}},\theta_{\mathrm{e}})$ and $Q_{\mathrm{e}}(\theta_{\mathrm{a}},\theta_{\mathrm{e}})$ using finite two-dimensional Fourier series as follows:
\begin{align}
\label{eq:phase_coupling_fourier_annual}
Q_{\mathrm{a}} (\theta_{\mathrm{a}}, \theta_{\mathrm{e}})
&= c_{\boldsymbol{0}}^{\mathrm{a}} + \sum_{\boldsymbol{m} \in \mathbb{M}}
c_{\boldsymbol{m}}^{\mathrm{a}} h_{\boldsymbol{m}} (\theta_{\mathrm{a}}, \theta_{\mathrm{e}}),
\\
\label{eq:phase_coupling_fourier_enso}
Q_{\mathrm{e}} (\theta_{\mathrm{a}}, \theta_{\mathrm{e}})
&= c_{\boldsymbol{0}}^{\mathrm{e}} + \sum_{\boldsymbol{m} \in \mathbb{M}}
c_{\boldsymbol{m}}^{\mathrm{e}} h_{\boldsymbol{m}} (\theta_{\mathrm{a}}, \theta_{\mathrm{e}}),
\end{align}
where $\boldsymbol{m}=(m_{\mathrm{a}},m_{\mathrm{e}})$ specifies the Fourier orders in $\theta_{\mathrm{a}}$ and $\theta_{\mathrm{e}}$, respectively.
The set $\mathbb{M}$ is defined by
\begin{align}
    \mathbb{M}
    := \left\{
        \boldsymbol m =(m_{\mathrm{a}},m_{\mathrm{e}}) \in\mathbb{Z}^2\setminus\{\boldsymbol{0}\} \;\middle|\;
        |m_{\mathrm{a}}|\le M,\,
        |m_{\mathrm{e}}|\le M
    \right\}.
\end{align}
The complex Fourier basis function corresponding to the Fourier index vector $\boldsymbol{m}$ is defined by
\begin{align}
    \label{eq:basisfunction_h}
    & h_{\boldsymbol{m}} (\theta_{\mathrm{a}}, \theta_{\mathrm{e}}) 
    := \exp[\mathrm{i}(m_{\mathrm{a}}\theta_{\mathrm{a}} + m_{\mathrm{e}}\theta_{\mathrm{e}})].
\end{align}
The constant term $c_{\boldsymbol 0}^{\mathrm p}$ ($\mathrm p\in\{\mathrm a,\mathrm e\}$) represents the effective frequency of the corresponding cycle.
We assume $c_{\boldsymbol 0}^{\mathrm p}=O(1)$ and $c_{\boldsymbol m}^{\mathrm p}h_{\boldsymbol m}(\theta_{\mathrm a},\theta_{\mathrm e})=O(\varepsilon)$ for all $\boldsymbol m\in\mathbb M$, where $\varepsilon\ll1$.
This assumption can be interpreted to mean that the phase-dependent modulation of the phase velocity is sufficiently small compared with the corresponding effective frequency.
We also assume that each entry of the noise covariance matrix ${\rm E}$ is $O(\varepsilon)$, corresponding to weak noise \citep{kuramoto_chemical_1984}.

\par
From the phase time-series data, we estimate the parameters $c_{\boldsymbol{m}}^{\mathrm{a}}$, $c_{\boldsymbol{m}}^{\mathrm{e}}$, and $\mathrm{E}$ in the stochastic phase equations.
There are $R=(2M+1)^2$ Fourier coefficients per phase equation, and thus the total number of parameters scales as $O(M^2)$.
The Bayesian inference method used to estimate these parameters is described in Appendix~\ref{appendix_B}.

\subsection{Averaging method}
\setsubsectionlabel{subsec:averaging_method}

\par
Although Eqs.~(\ref{eq:phase_equation_annual}) and (\ref{eq:phase_equation_enso}) describe modulation of the instantaneous frequencies as a function of the two phase variables, it is difficult to interpret the synchronization properties between the annual and ENSO cycles from these equations.
We therefore apply time averaging to derive an averaged form that clarifies synchronization properties between the two cycles \citep{kuramoto_chemical_1984,nakao_phase_2016,pietras_network_2019}.

\par
For positive integers $m$ and $n$, we introduce the phase-difference variable
\begin{align}
    \label{eq:psi_mn}
    \psi_{mn}:=\theta_{\mathrm{a}}-\frac{m}{n}\theta_{\mathrm{e}}, 
\end{align}
whose dynamics characterize the synchronization properties between the annual and ENSO cycles.
For instance, $\dot{\psi}_{mn}=0$ for all time corresponds to $m:n$ phase locking with a constant value of $\psi_{mn}$.
Averaging yields an approximate equation for the evolution of the phase difference.

\par
We denote the effective frequencies associated with the constant Fourier terms by
\begin{align}
    \label{eq:const_omega_a}
    & \omega_{\mathrm{a}} = c_{\boldsymbol{0}}^{\mathrm{a}}, \\
    \label{eq:const_omega_e}
    & \omega_{\mathrm{e}} = c_{\boldsymbol{0}}^{\mathrm{e}}.
\end{align}
We assume the near-resonance condition
\begin{align}
    \label{eq:near_resonance_condition}
    n\omega_{\mathrm{a}}-m\omega_{\mathrm{e}}
    &=O(\varepsilon).
\end{align}
Under this condition, the phase difference $\psi_{mn}$ varies slowly enough to be treated as approximately constant over an interval spanning $m$ annual cycles and $n$ ENSO cycles~\citep{pikovsky_synchronization_2001}.

\par
We apply the averaging method to $Q_{\mathrm a}(\theta_{\mathrm a},\theta_{\mathrm e})$ and $Q_{\mathrm e}(\theta_{\mathrm a},\theta_{\mathrm e})$ over $n$ ENSO cycles while holding the slow phase-difference variable $\psi_{mn}$ fixed~\citep{pikovsky_synchronization_2001}:
\begin{align}
    \label{eq:overline_Qa}
    \overline{Q}_{\mathrm{a}}(\psi_{mn})
    &:=
    \frac{1}{2\pi n}
    \int_0^{2\pi n}
    Q_{\mathrm{a}}
    \!\left(
        \psi_{mn}+\frac{m}{n}\theta,
        \theta
    \right)
    \,d\theta,
    \\
    \label{eq:overline_Qe}
    \overline{Q}_{\mathrm{e}}(\psi_{mn})
    &:=
    \frac{1}{2\pi n}
    \int_0^{2\pi n}
    Q_{\mathrm{e}}
    \!\left(
        \psi_{mn}+\frac{m}{n}\theta,
        \theta
    \right)
    \,d\theta.
\end{align}
Averaging retains a Fourier term if and only if its index vector $\boldsymbol{m} \in \mathbb{M}$ satisfies the resonance condition
\begin{align}
    \label{eq:resonance_condition}
    m m_{\mathrm a}+n m_{\mathrm e}=0.
\end{align}
On the basis of the resonance condition, we define the set of admissible positive Fourier orders by
\begin{align}
    \label{eq:admissible_fourier_orders}
    \mathcal K_{mn}
    :=\left\{
        k\in\mathbb Z_{>0}
        \;\middle|\;
        \left(k,-\frac{m}{n}k\right)\in\mathbb M
    \right\}.
\end{align}
For each $k\in\mathcal K_{mn}$, we define the corresponding Fourier index vector by
\begin{align}
    \label{eq:resonant_index_vector}
    \boldsymbol m_k(m,n)
    :=\left(k,-\frac{m}{n}k\right)\in\mathbb M.
\end{align}

\par
Because the index vector $\boldsymbol m_k(m,n)$ satisfies the resonance condition, averaging retains its Fourier term as the $k$th harmonic in $\psi_{mn}$.
Specifically, substituting Eqs.~(\ref{eq:phase_coupling_fourier_annual}) and (\ref{eq:phase_coupling_fourier_enso}) into Eqs.~(\ref{eq:overline_Qa}) and (\ref{eq:overline_Qe}) gives the averaged functions
\begin{align}
    \label{eq:averaged_Q_fourier_annual}
    \overline Q_{\mathrm a}(\psi_{mn})
    &=
    c_{\boldsymbol 0}^{\mathrm a}
    +2\operatorname{Re}
    \sum_{k\in\mathcal K_{mn}}
    c_{\boldsymbol m_k(m,n)}^{\mathrm a}
    \exp\!\left(\mathrm i k\psi_{mn}\right),
    \\
    \label{eq:averaged_Q_fourier_enso}
    \overline Q_{\mathrm e}(\psi_{mn})
    &=
    c_{\boldsymbol 0}^{\mathrm e}
    +2\operatorname{Re}
    \sum_{k\in\mathcal K_{mn}}
    c_{\boldsymbol m_k(m,n)}^{\mathrm e}
    \exp\!\left(\mathrm i k\psi_{mn}\right),
\end{align}
where the zero mode is also retained after averaging, yielding the constant terms $c_{\boldsymbol0}^{\mathrm a}$ and $c_{\boldsymbol0}^{\mathrm e}$.

\par
We define these phase-dependent terms as the effective phase coupling functions $\Gamma_{\mathrm a}(\psi_{mn})$ and $\Gamma_{\mathrm e}(\psi_{mn})$:
\begin{align}
    \label{eq:averaging_Gamma_a}
    \Gamma_{\mathrm a}(\psi_{mn})
    & =
    2\operatorname{Re}
    \sum_{k\in\mathcal K_{mn}}
    c_{\boldsymbol m_k(m,n)}^{\mathrm a}
    \exp\!\left(\mathrm i k\psi_{mn}\right)
    \nonumber \\
    &=
    \sum_{k\in\mathcal K_{mn}}
    \left[
        a_k^{\mathrm a}\cos(k\psi_{mn})
        +b_k^{\mathrm a}\sin(k\psi_{mn})
    \right],
    \\
    \label{eq:averaging_Gamma_e}
    \Gamma_{\mathrm e}(\psi_{mn})
    & =
    2\operatorname{Re}
    \sum_{k\in\mathcal K_{mn}}
    c_{\boldsymbol m_k(m,n)}^{\mathrm e}
    \exp\!\left(\mathrm i k\psi_{mn}\right)
    \nonumber \\
    &=
    \sum_{k\in\mathcal K_{mn}}
    \left[
        a_k^{\mathrm e}\cos(k\psi_{mn})
        +b_k^{\mathrm e}\sin(k\psi_{mn})
    \right].
\end{align}
For $k\in\mathcal K_{mn}$, the real Fourier coefficients are given by
\begin{align}
    \label{eq:resonant_coefficient_a_cos}
    a_k^{\mathrm a}
    &=2\operatorname{Re}
    c_{\boldsymbol m_k(m,n)}^{\mathrm a},
    \\
    \label{eq:resonant_coefficient_e_cos}
    a_k^{\mathrm e}
    &=2\operatorname{Re}
    c_{\boldsymbol m_k(m,n)}^{\mathrm e},
    \\
    \label{eq:resonant_coefficient_a_sin}
    b_k^{\mathrm a}
    &=-2\operatorname{Im}
    c_{\boldsymbol m_k(m,n)}^{\mathrm a},
    \\
    \label{eq:resonant_coefficient_e_sin}
    b_k^{\mathrm e}
    &=-2\operatorname{Im}
    c_{\boldsymbol m_k(m,n)}^{\mathrm e}.
\end{align}

\par 
Applying the averaging procedure in Eqs.~(\ref{eq:const_omega_a})--(\ref{eq:resonant_coefficient_e_sin}) yields the following approximate phase equations:
\begin{align}
    \label{eq:averaging_correspondence_annual}
    \dot{\theta}_\mathrm{a}
    &\simeq \omega_\mathrm{a} + \Gamma_{\mathrm{a}}(\psi_{mn}),
    \\
    \label{eq:averaging_correspondence_enso}
    \dot{\theta}_\mathrm{e}
    &\simeq \omega_\mathrm{e} + \Gamma_{\mathrm{e}}(\psi_{mn}).
\end{align}
Thus, in the averaged form, the modulation depends only on $\psi_{mn}$ rather than on $\theta_{\mathrm a}$ and $\theta_{\mathrm e}$.
Combining Eqs.~(\ref{eq:averaging_correspondence_annual}) and (\ref{eq:averaging_correspondence_enso}), we obtain the following equation for the time evolution of the phase difference $\psi_{mn}$:
\begin{align}
    \label{eq:phase_difference_dynamics_withoutnoise}
    \dot{\psi}_{mn} &\simeq \omega_{\mathrm{a}} -\frac{m}{n}\omega_{\mathrm{e}}
    +\Gamma_{\mathrm{a}}(\psi_{mn}) -\frac{m}{n}\Gamma_{\mathrm{e}}(\psi_{mn}).
\end{align}
This equation can be used to examine the properties of $m:n$ synchronization.
In Eqs.~(\ref{eq:averaging_correspondence_annual})--(\ref{eq:phase_difference_dynamics_withoutnoise}), we omit the noise terms and present only the deterministic parts obtained by averaging.

\par
The higher-order harmonics in Eqs.~(\ref{eq:averaged_Q_fourier_annual}) and (\ref{eq:averaged_Q_fourier_enso}) are constrained by both the Fourier truncation order $M$ used to define $\mathbb M$ and the ratio $m:n$.
From the definitions of $\mathcal K_{mn}$ and $\mathbb M$, the Fourier orders $k\in\mathcal K_{mn}$ satisfy
\begin{align}
    \label{eq:k_cutoff}
    k\leq\min\!\left(M,\frac{nM}{m}\right).
\end{align}
Thus, no harmonics above order $M$ can appear in Eqs.~(\ref{eq:averaged_Q_fourier_annual}) and (\ref{eq:averaged_Q_fourier_enso}).
For $m>n$, the upper bound is further tightened to $(n/m)M$.

\subsection{Direct reconstruction of the averaged form of the phase equations}
\setsubsectionlabel{subsec:direct_averaged_phase_inference}

\par
As an alternative approach to the averaging method described in the previous subsection, we directly reconstruct the averaged form of the phase equations, Eqs.~(\ref{eq:averaging_correspondence_annual}) and (\ref{eq:averaging_correspondence_enso}), from the phase time series~\citep{ota_direct_2014,onojima_dynamical_2018,ota_interaction_2020,arai_extracting_2022,furukawa_bayesian_2025}.
This approach allows us to capture higher-order harmonics beyond the cutoff specified in Eq.~(\ref{eq:k_cutoff}).

\par
We consider the averaged form of the stochastic phase equations:
\begin{align}
    \label{eq:averaged_phase_equation_annual}
    & \dot{\theta}_{\mathrm{a}}(t) 
    = \omega_{\mathrm{a}} 
    + \Gamma_{\mathrm{a}}(\psi_{mn}) 
    + \zeta_{\mathrm{a}}(t), \\
    \label{eq:averaged_phase_equation_enso}
    & \dot{\theta}_{\mathrm{e}}(t) 
    = \omega_{\mathrm{e}} 
    + \Gamma_{\mathrm{e}}(\psi_{mn}) 
    + \zeta_{\mathrm{e}}(t),
\end{align}
where Gaussian noise terms $\zeta_{\mathrm{a}}(t)$ and $\zeta_{\mathrm{e}}(t)$ are added to Eqs.~(\ref{eq:averaging_correspondence_annual}) and (\ref{eq:averaging_correspondence_enso}).
We truncate the phase coupling functions at Fourier order $M$, allowing higher-order harmonics beyond the cutoff specified in Eq.~(\ref{eq:k_cutoff}).
\begin{align}
    \label{eq:direct_coupling_fourier}
    \Gamma_{\mathrm a}(\psi_{mn})
    &=
    \sum_{k=1}^{M}
    \left[
        a_k^{\mathrm a}\cos(k\psi_{mn})
        +b_k^{\mathrm a}\sin(k\psi_{mn})
    \right],
    \\
    \label{eq:direct_coupling_fourier_enso}
    \Gamma_{\mathrm e}(\psi_{mn})
    &=
    \sum_{k=1}^{M}
    \left[
        a_k^{\mathrm e}\cos(k\psi_{mn})
        +b_k^{\mathrm e}\sin(k\psi_{mn})
    \right].
\end{align}
The mean and intensity of the noise are specified as follows:
\begin{align}
    \label{eq:averaged_noise_mean}
    \left\langle
        \zeta_{\mathrm{p}}(t)
    \right\rangle
    &=
    0
    \qquad
    (\mathrm{p}\in\{\mathrm{a},\mathrm{e}\}),
    \\
    \label{eq:averaged_noise_intensity}
    \left\langle
        \zeta_{\mathrm{p}}(t)
        \zeta_{\mathrm{q}}(t')
    \right\rangle
    &=
    \sigma_{\mathrm{p}}^2
    \delta_{\mathrm{p}\mathrm{q}}
    \delta(t-t')
    \qquad
    (\mathrm{p},\mathrm{q}\in\{\mathrm{a},\mathrm{e}\}).
\end{align}
We assume that $\zeta_{\mathrm{a}}$ and $\zeta_{\mathrm{e}}$ are independent.
This simplification is motivated by the weak correlation between the phase noise terms of the annual and ENSO cycles, as shown subsequently in Sec.~\ref{subsec:phase_dynamics_results}.
From the phase time-series data, we estimate the parameters $\omega_{\mathrm{p}}$, $\{a_k^{\mathrm{p}},b_k^{\mathrm{p}}\}_{k=1}^{M}$, and $\sigma_{\mathrm{p}}^2$.
There are $R':=2M+1$ Fourier coefficients per phase equation, and thus the total number of parameters scales as $O(M)$.
The Bayesian inference method used to estimate these parameters is described in Appendix~\ref{appendix_C}.

\par
Once Eqs.~(\ref{eq:averaged_phase_equation_annual}) and (\ref{eq:averaged_phase_equation_enso}) are reconstructed, the following equation for the time evolution of $\psi_{mn}$ is derived:
\begin{align}
    \label{eq:phase_difference_dynamics_withnoise}
    \dot{\psi}_{mn} &= \omega_{\mathrm{a}} -\frac{m}{n}\omega_{\mathrm{e}} 
    +\Gamma_{\mathrm{a}}(\psi_{mn}) -\frac{m}{n}\Gamma_{\mathrm{e}}(\psi_{mn}) + \zeta(t), \\
    \label{eq:phase_difference_noise}
    \zeta(t)
    &:=
    \zeta_{\mathrm{a}}(t)
    -
    \frac{m}{n}\zeta_{\mathrm{e}}(t).
\end{align}
As in Eq.~(\ref{eq:phase_difference_dynamics_withoutnoise}), the deterministic part of Eq.~(\ref{eq:phase_difference_dynamics_withnoise}) can be used to assess $m:n$ synchronization.

\par
Reconstructing the averaged form directly requires estimating fewer parameters than first reconstructing the two-variable form, $Q_{\mathrm p}(\theta_{\mathrm a},\theta_{\mathrm e})$, and then averaging it.
To obtain the averaged form with harmonics up to order $M$, direct reconstruction requires estimating $O(M)$ parameters per phase equation, whereas reconstructing the two-variable form before averaging requires $O(M^2)$.

\section{Results}
\label{sec:results}

\par
In this section, we present the results of our analysis.
In Sec.~\ref{subsec:koopman_results}, we present the phase time series computed with the data-driven Koopman operator approach.
We then present the reconstructed phase dynamics in Sec.~\ref{subsec:phase_dynamics_results}.
In Sec.~\ref{subsec:averaging_results}, we present their averaged form derived by applying the averaging method.
In Sec.~\ref{subsec:direct_inference_results}, we present the averaged phase equations reconstructed directly from the phase time series.

\paragraph*{SST data}
We used monthly mean SST data from ERSSTv5 \citep{huang_ersstv5_2017}.
The analysis period extended from January 1970 to December 2025, with a sampling interval of $\Delta t = 1/12~[\mathrm{year}]$.
We used SST fields over the Indo-Pacific domain spanning $28^{\circ}\mathrm{E}$--$290^{\circ}\mathrm{E}$ and $60^{\circ}\mathrm{S}$--$20^{\circ}\mathrm{N}$, with a spatial resolution of $2^{\circ}$.
No detrending was applied to the SST data.
(We also analyzed HadISST \citep{rayner_hadisst_2003} and COBE-SST2 \citep{hirahara_cobesst2_2014} with results qualitatively similar to those in Sec.~\ref{sec:results}; these results are not shown.)

\subsection{Phases of the annual and ENSO cycles obtained from Koopman eigenfunctions}
\setsubsectionlabel{subsec:koopman_results}

\par
Using the data-driven Koopman operator approach, we obtained time series of the eigenfunctions.
The resulting eigenfunction time series extended from March 1972 to November 2023.
Among the eigenfunctions obtained, we selected $g_{\mathrm{a}}(\omega_\ell)$ and $g_{\mathrm{e}}(\omega_\ell)$ representing the annual and ENSO cycles, respectively, based on the eigenfrequencies in Eqs.~(\ref{eq:koopman_eigenfrequency_annual}) and (\ref{eq:koopman_eigenfrequency_enso}).
The corresponding eigenfrequencies were
\begin{align}
    \label{eq:koopman_eigenfrequency_annual_result}
    \frac{\operatorname{Im}\nu_{\mathrm{a}}}{2\pi}
    &= 0.9912 \quad [\mathrm{cycle / year}], \\
    \label{eq:koopman_eigenfrequency_enso_result}
    \frac{\operatorname{Im}\nu_{\mathrm{e}}}{2\pi}
    &= 0.2439 \quad [\mathrm{cycle / year}].
\end{align}

\par
\begin{figure*}[t]
  \noindent\includegraphics[width=\textwidth,angle=0]{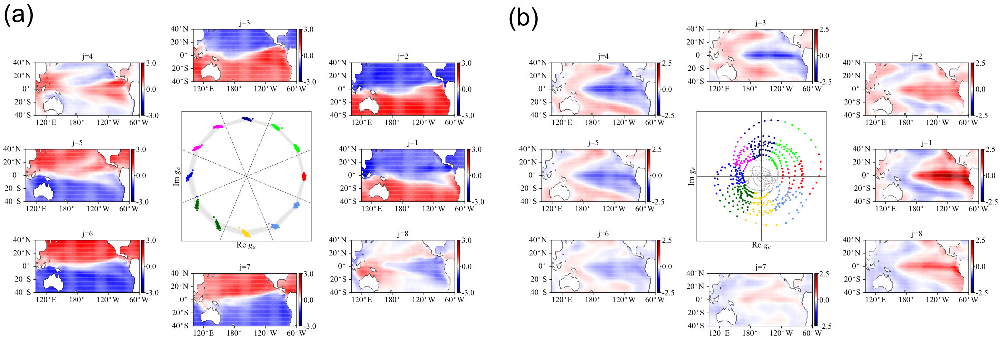}\\
  \caption{
        Eigenfunctions and SST phase composites of the annual and ENSO cycles.
        (a)~Eigenfunction $g_{\mathrm a}(\omega_\ell)$ and phase composites of SST fields.
        The central plot shows $g_{\mathrm a}(\omega_\ell)$ in the complex plane.
        The colors of the data points indicate the stage to which each sample belongs.
        The surrounding maps show phase composites of SST fields standardized at each grid point for stages $j=1,\ldots,8$.
        Stages $j=1$ and $j=5$ correspond to January and July, respectively.
        (b)~Same as panel (a) but for the eigenfunction $g_{\mathrm e}(\omega_\ell)$ and phase composites of monthly SSTA fields.
        Stages $j=8,1,2$ are classified as El Ni\~{n}o stages and stages $j=3,4,5$ as La Ni\~{n}a stages.
        In both panels, data points with $R_{\mathrm e}(t_\ell)<\tilde{R}_{\mathrm e}$ are shown as small gray circles, indicating that the ENSO cycle is considered inactive. 
        In panel (a), points from the same calendar month are tightly clustered, so the gray circles are hidden behind the clusters of colored circles. 
        In panel (b), the gray circles lie within the region $R_{\mathrm e}(t_\ell)<\tilde{R}_{\mathrm e}$.
    }
  \label{fig:fig1}
\end{figure*}

The phases of the annual and ENSO cycles were computed from Eqs.~(\ref{eq:phase_a_from_eigenfunction}) and (\ref{eq:phase_e_from_eigenfunction}), respectively.
We visualized the relationship between phase and SST by constructing phase composites as described in Eqs.~(\ref{eq:phase_bin})--(\ref{eq:phase_composite_enso}).
We divided each cycle into $S=8$ stages.
We used the following amplitude threshold:
\begin{align}
    \label{eq:R_threshold_30th}
    \tilde{R}_{\mathrm{e}} = \tilde{R}_{30\,\mathrm{th}},
\end{align}
where $\tilde{R}_{a\,\mathrm{th}}$ denotes the $a$th percentile of $\{R_{\mathrm{e}}(t_\ell)\}_\ell$.
Figures~\ref{fig:fig1}(a) and \ref{fig:fig1}(b) show $g_{\mathrm a}(\omega_\ell)$ and $g_{\mathrm e}(\omega_\ell)$, respectively, with surrounding maps showing phase composites for stages $j=1,2,\ldots,8$.

\par
Figure~\ref{fig:fig1}(a) shows that the phase of the annual cycle takes nearly the same value for a given calendar month across years, accurately capturing the phase progression of the annual cycle.
Stages $j=1$ and $j=5$ correspond to January and July, respectively.
Figure~\ref{fig:fig1}(b) shows that El Ni\~{n}o and La Ni\~{n}a do not occupy directly opposite phases in the ENSO cycle represented by the Koopman eigenfunction, consistent with the findings of \citet{froyland_spectral_2021}.
La Ni\~{n}a appears earlier than halfway through the cycle.
We classified stages $j=8,1,2$ as El Ni\~{n}o stages and stages $j=3,4,5$ as La Ni\~{n}a stages.
The monthly SSTA phase composites averaged over the Ni\~{n}o 3.4 region exceeded $0.5~^{\circ}\mathrm{C}$ for the El Ni\~{n}o stages and fell below $-0.5~^{\circ}\mathrm{C}$ for the La Ni\~{n}a stages.

\par
As mentioned in Sec.~\ref{subsec:operator_theoretic_formalism}, samples with $R_{\mathrm e}(t_\ell) < \tilde{R}_{\mathrm e}$ were classified as inactive and can be interpreted as states near the phase singularity of the ENSO cycle.
These samples correspond to the points shown in gray in Figs.~\ref{fig:fig1}(a) and \ref{fig:fig1}(b).
Phase data associated with these inactive states were excluded from the datasets used for Bayesian inference based on the models specified in Secs.~\ref{subsec:phase_dynamics_reconstruction} and \ref{subsec:direct_averaged_phase_inference}.
Procedures for constructing these datasets are described in Appendices~\ref{appendix_B} and \ref{appendix_C}.

\subsection{Seasonal modulation of the phase dynamics of the ENSO cycle}
\setsubsectionlabel{subsec:phase_dynamics_results}

\par
\begin{figure*}[t]
  \noindent\includegraphics[width=\textwidth,angle=0]{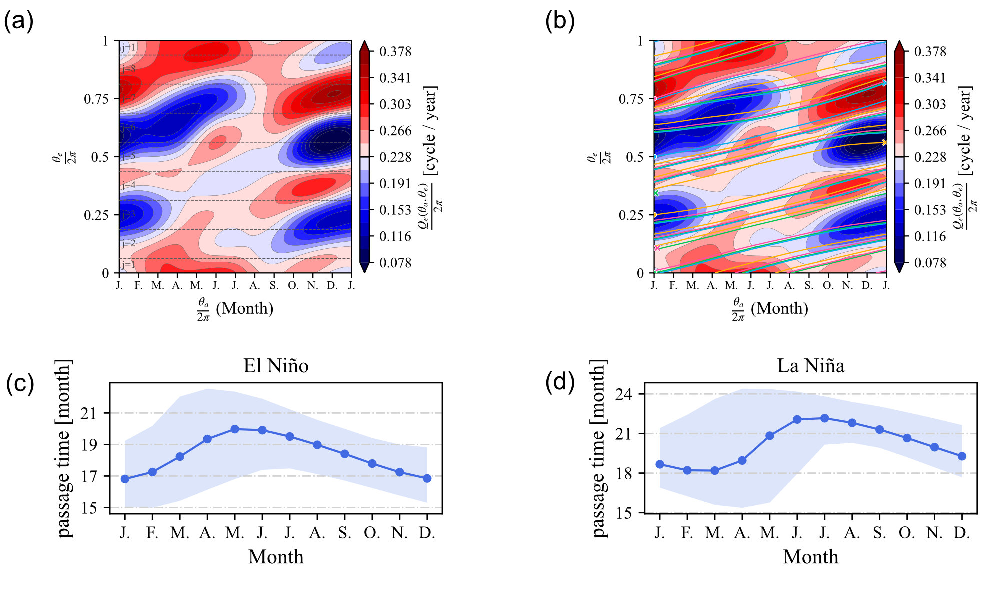}\\
    \caption{
        Reconstructed phase equation and quantification of the durations of the El Ni\~{n}o and La Ni\~{n}a stages.
        (a)~Reconstructed function $Q_{\mathrm e}(\theta_{\mathrm a},\theta_{\mathrm e})$ in the phase equation for the ENSO cycle.
        Since $Q_{\mathrm a}(\theta_{\mathrm a},\theta_{\mathrm e})/(2\pi)\simeq1~[\mathrm{cycle/year}]$, the annual phase advances nearly uniformly and approximately corresponds to the calendar month.
        Month $\mu$ corresponds approximately to $\theta_{\mathrm a}/(2\pi)=(\mu-1)/12$, where $\mu=1,\ldots,12$ denotes January through December.
        Horizontal dashed lines indicate the boundaries between the ENSO stages $j=1,\ldots,8$ defined in Eq.~(\ref{eq:phase_bin}).
        (b)~Sample deterministic trajectories in the $(\theta_{\mathrm a},\theta_{\mathrm e})$ plane.
        The trajectories advance horizontally by approximately one cycle per year, whereas their vertical motion speeds up and slows down according to $Q_{\mathrm e}(\theta_{\mathrm a},\theta_{\mathrm e})$.
        (c)~First passage times through the El Ni\~{n}o stages, estimated from Langevin simulations of the reconstructed stochastic phase equations.
        The horizontal axis indicates the starting month $\mu$, which determines the initial condition $\theta_{\mathrm a}(0)=\theta_\mu$.
        For each starting month $\mu=1,\ldots,12$, 5000 simulations were performed.
        The solid line shows the mean, while the shading indicates the 2.5th--97.5th percentile range.
        The first passage times through the El Ni\~{n}o stages are interpreted as the durations of these stages.
        (d)~Same as panel (c) but for the La Ni\~{n}a stages.
    }
    \label{fig:fig2}
\end{figure*}

From the phase time-series data of the annual and ENSO cycles, we reconstructed the functions $Q_\mathrm{a}(\theta_{\mathrm{a}},\theta_{\mathrm{e}})$ and $Q_\mathrm{e}(\theta_{\mathrm{a}},\theta_{\mathrm{e}})$ with the Fourier truncation order set to $M=3$.
Figure~\ref{fig:fig2}(a) shows $Q_\mathrm{e}(\theta_{\mathrm{a}},\theta_{\mathrm{e}})$.
The reconstructed function $Q_{\mathrm{e}}(\theta_{\mathrm{a}},\theta_{\mathrm{e}})$ exhibits distinct bands sloping upward to the right in the $(\theta_{\mathrm{a}},\theta_{\mathrm{e}})$ plane.
The contrast between high and low values is particularly pronounced during boreal winter and spring.
We also reconstructed the phase equation for the annual cycle, but do not show it here because $\frac{Q_{\mathrm{a}}(\theta_{\mathrm{a}},\theta_{\mathrm{e}})}{2\pi}\simeq1~\mathrm{[cycle/year]}$.
The results described in this paragraph and shown in Fig.~\ref{fig:fig2}(a) remained qualitatively unchanged when the Fourier truncation order was increased to $M=5$.

\par 
The estimated noise covariance matrix is
\begin{align}
    \label{eq:estimated_noise_intensity_matrix}
    \frac{\mathrm E}{(2\pi)^2}
    &=
    \begin{pmatrix}
    1.712 \times 10^{-5} & -6.956 \times 10^{-6}\\
    -6.956 \times 10^{-6} & 3.864 \times 10^{-4}
    \end{pmatrix}
    \quad [\mathrm{cycle}^2/\mathrm{year}].
\end{align}
This matrix yields a correlation coefficient of approximately $-0.086$, indicating weak correlation between the noise terms $\eta_{\mathrm{a}}(t)$ and $\eta_{\mathrm{e}}(t)$ in Eqs.~(\ref{eq:phase_equation_annual}) and (\ref{eq:phase_equation_enso}).
On the basis of this result, we simplified direct reconstruction of the averaged phase equations by assuming independent noise terms.

\par
To visualize phase dynamics, we numerically integrated the reconstructed phase equations without noise and overlaid the trajectories on a plot of $Q_{\mathrm{e}}(\theta_{\mathrm{a}},\theta_{\mathrm{e}})$ in Fig.~\ref{fig:fig2}(b).
Since $\frac{Q_{\mathrm{a}}(\theta_{\mathrm{a}},\theta_{\mathrm{e}})}{2\pi}\simeq1~\mathrm{[cycle/year]}$, $\theta_{\mathrm{a}}(t)$ advances nearly uniformly at approximately one cycle per year.
The instantaneous frequency of the ENSO cycle is modulated according to $Q_{\mathrm{e}}(\theta_{\mathrm{a}},\theta_{\mathrm{e}})$.
The dependence of this modulation on $\theta_{\mathrm{a}}$ reflects how the ENSO phase responds to coupling with the annual cycle.
Thus, the state $(\theta_{\mathrm{a}},\theta_{\mathrm{e}})$ in Fig.~\ref{fig:fig2}(b) moves horizontally at a nearly constant speed, whereas its vertical speed varies according to $Q_{\mathrm{e}}(\theta_{\mathrm{a}},\theta_{\mathrm{e}})$.

\par
As described in Sec.~\ref{subsec:koopman_results}, the intervals of $\theta_{\mathrm{e}}$ associated with El Ni\~{n}o and La Ni\~{n}a correspond to stages $j=8,1,2$ and $j=3,4,5$, respectively.
Thus, for El Ni\~{n}o, we define the initial and target values of $\theta_{\mathrm{e}}$ as $\theta_{\mathrm{EN}}^{\mathrm{start}}=13\pi/8$ and $\theta_{\mathrm{EN}}^{\mathrm{end}}=2\pi+3\pi/8$, respectively.
For La Ni\~{n}a, we define the initial and target values of $\theta_{\mathrm{e}}$ as $\theta_{\mathrm{LN}}^{\mathrm{start}}=3\pi/8$ and $\theta_{\mathrm{LN}}^{\mathrm{end}}=9\pi/8$, respectively.
For the annual cycle, we define the phase for month $\mu$ as $\theta_\mu=\frac{2\pi(\mu-1)}{12}$, where $\mu=1,\ldots,12$ corresponds to January through December.

\par 
We calculated the durations of the El Ni\~{n}o and La Ni\~{n}a stages starting in month $\mu$ as first passage times through numerical integration of the reconstructed stochastic phase equations~\citep{redner_guide_2001,kloeden_numerical_1992}.
For an El Ni\~{n}o phase starting in month $\mu$, we use the initial condition
\begin{align}
    \label{eq:el_nino_initial_condition}
    \left(\theta_{\mathrm{a}}(0),\,\theta_{\mathrm{e}}(0)\right)
    &=
    \left(\theta_\mu,\,\theta_{\mathrm{EN}}^{\mathrm{start}}\right).
\end{align}
The first passage time $\tau_{\mu,\mathrm{EN}}$ is defined as the smallest positive time satisfying $\theta_{\mathrm{e}}(\tau_{\mu,\mathrm{EN}})=\theta_{\mathrm{EN}}^{\mathrm{end}}$.
Similarly, for a La Ni\~{n}a phase starting in month $\mu$, we use the initial condition
\begin{align}
    \label{eq:la_nina_initial_condition}
    \left(\theta_{\mathrm{a}}(0),\,\theta_{\mathrm{e}}(0)\right)
    &=
    \left(\theta_\mu,\,\theta_{\mathrm{LN}}^{\mathrm{start}}\right).
\end{align}
The first passage time $\tau_{\mu,\mathrm{LN}}$ is defined as the smallest positive time satisfying $\theta_{\mathrm{e}}(\tau_{\mu,\mathrm{LN}})=\theta_{\mathrm{LN}}^{\mathrm{end}}$.

\par
For each starting month $\mu$, we numerically integrated the reconstructed stochastic phase equations 5000 times for each of the El Ni\~{n}o and La Ni\~{n}a initial conditions.
This yielded 5000 samples of each of $\tau_{\mu,\mathrm{EN}}$ and $\tau_{\mu,\mathrm{LN}}$ for $\mu=1,2,\ldots,12$.
The resulting statistics are shown in Figs.~\ref{fig:fig2}(c) and \ref{fig:fig2}(d), respectively.
For the El Ni\~{n}o stages, the mean duration is longest for starts from May to July, at approximately 20 months.
It is shorter for starts from November to February, at approximately 17 months.
For the La Ni\~{n}a stages, the mean duration is longest for starts from June to August, at approximately 22 months.
It is shorter for starts from January to April, at approximately 18--19 months.

\par
The variability in duration also shows a seasonal dependence on the starting month.
The shaded regions in Figs.~\ref{fig:fig2}(c) and \ref{fig:fig2}(d) indicate the 2.5th--97.5th percentile ranges.
For both the El Ni\~{n}o and La Ni\~{n}a stages, these ranges are wider for starts in boreal spring and narrower for starts in late summer and autumn.
Both the mean and variability of the durations show seasonal dependence, although they peak for different starting months.

\par
Results for the amplitude thresholds $\tilde{R}_{10\,\mathrm{th}}, \tilde{R}_{20\,\mathrm{th}}, \ldots, \tilde{R}_{50\,\mathrm{th}}$ appear in Appendix~\ref{appendix_D}.
Qualitatively similar results were obtained for all these thresholds.

\subsection{Averaged form of the phase equations indicating statistical $3:1$ phase locking}
\setsubsectionlabel{subsec:averaging_results}

\par
To investigate annual--ENSO synchronization, we applied the averaging method described in Sec.~\ref{subsec:averaging_method} to the previously reconstructed two-variable functions $Q_{\mathrm a}(\theta_{\mathrm{a}},\theta_{\mathrm{e}})$ and $Q_{\mathrm e}(\theta_{\mathrm{a}},\theta_{\mathrm{e}})$.

\par
\begin{figure}[t]
  \centering
  \includegraphics[width=19pc,angle=0]{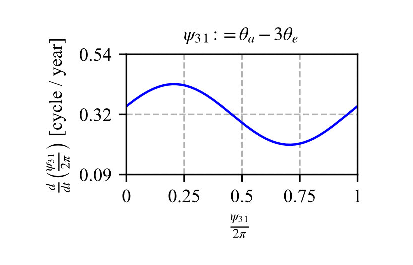}
  \caption{
        Equation~(\ref{eq:phase_difference_dynamics_withoutnoise}) for the time evolution of the phase difference $\psi_{3\,1}$.
        This equation is obtained by applying the averaging method to the reconstructed $Q_{\mathrm a}(\theta_{\mathrm a},\theta_{\mathrm e})$ and $Q_{\mathrm e}(\theta_{\mathrm a},\theta_{\mathrm e})$ as described in Sec.~\ref{subsec:averaging_method}.
        }
\label{fig:fig3}
\end{figure}

Figure~\ref{fig:fig3} shows the equation for the time evolution of the phase difference $\psi_{m\,n}$, Eq.~(\ref{eq:phase_difference_dynamics_withoutnoise}), with the frequency ratio $m:n=3:1$.
In this equation, the constant term is $\omega_{\mathrm a}-3\omega_{\mathrm e}$.
The $\psi_{3\,1}$-dependent modulation of $\dot{\psi}_{3\,1}$ is mainly determined by $\Gamma_{\mathrm e}(\psi_{3\,1})$, since $(\omega_{\mathrm a}+\Gamma_{\mathrm a}(\psi_{3\,1}))/(2\pi)\simeq1~[\mathrm{cycle/year}]$.

\par
The equation for $\psi_{3\,1}$ has no stable fixed point and therefore does not support complete phase locking.
However, there is a region over which $\psi_{3\,1}$ advances much more slowly; the velocity corresponds to approximately 2.3 years per cycle in the fastest region and 4.9 years per cycle in the slowest region.
This suggests statistical $3:1$ phase locking.

\par 
Under the cutoff condition in Eq.~(\ref{eq:k_cutoff}) with $M=3$, the averaged form of the phase equations contains only the first Fourier harmonic for $m:n=2:1$ and $3:1$.
For $m:n=4:1$ and $5:1$, averaging removes even the first Fourier harmonic, leaving only constant terms.
To overcome this limitation, we directly reconstruct the averaged form of the phase equations from the phase time-series data in the next subsection.

\subsection{Direct reconstruction of the averaged form for assessing statistical $m:n$ phase locking}
\setsubsectionlabel{subsec:direct_inference_results}

\par
Using the method described in Sec.~\ref{subsec:direct_averaged_phase_inference}, we directly reconstructed the averaged form of the phase equations, Eqs.~(\ref{eq:averaged_phase_equation_annual}) and (\ref{eq:averaged_phase_equation_enso}), from the phase time-series data with the Fourier truncation order set to $M=3$.

\par
\begin{figure}[t]
  \centering
  \includegraphics[width=19pc,angle=0]{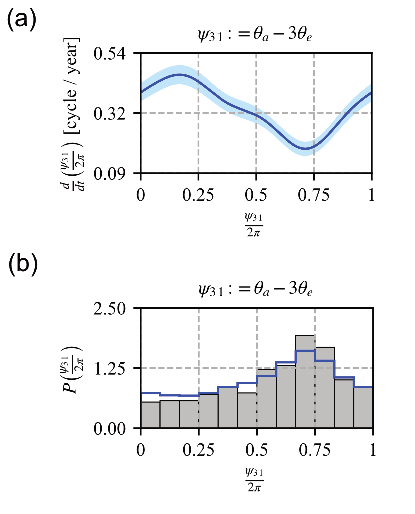}
  \caption{
        Results of the direct reconstruction of the phase equations in averaged form.
        (a)~
        Equation for the time evolution of the phase difference $\psi_{3\,1}$, directly reconstructed from the phase time-series data.
        The solid curve shows the deterministic part of the phase equation reconstructed by Bayesian inference, while the shading indicates the estimated standard deviation.
        (b)~
        Histograms of $\psi_{3\,1}$.
        The histogram obtained from the phase time-series data is shown in gray.
        Samples with $R_{\mathrm e} < \tilde{R}_{\mathrm e}$ are excluded from this histogram.
        The blue line shows the histogram obtained from a Langevin simulation of the reconstructed stochastic phase equations over 5000 years.
        }
    \label{fig:fig4}
\end{figure}

Figure~\ref{fig:fig4}(a) shows the deterministic part of Eq.~(\ref{eq:phase_difference_dynamics_withnoise}) with $m:n=3:1$, which describes the time evolution of the phase difference $\psi_{3\,1}$. 
Although the directly reconstructed equation includes harmonics up to order $M=3$, its deterministic part is qualitatively similar to that shown in Fig.~\ref{fig:fig3}, which contains only the first harmonic.
As in the results obtained using the averaging method, the $\psi_{3\,1}$-dependent modulation of the deterministic part is mainly determined by $\Gamma_{\mathrm e}(\psi_{3\,1})$, since $(\omega_{\mathrm a}+\Gamma_{\mathrm a}(\psi_{3\,1}))/(2\pi)\simeq1~[\mathrm{cycle/year}]$.

\par
Additionally, Fig.~\ref{fig:fig4}(b) compares histograms of $\{\psi_{3\,1}(t_\ell)\}_\ell$ obtained from the phase time-series data and from Langevin simulations of the reconstructed stochastic phase equations.
In the Langevin simulations of Eqs.~(\ref{eq:averaged_phase_equation_annual}) and (\ref{eq:averaged_phase_equation_enso}), we used the estimated noise intensities
\begin{align}
    \frac{\sigma_{\mathrm{a}}^2}{(2\pi)^2}
    &= 5.016\times10^{-5}
    \quad [\mathrm{cycle^2/year}], \\
    \frac{\sigma_{\mathrm{e}}^2}{(2\pi)^2}
    &= 5.372\times10^{-4}
    \quad [\mathrm{cycle^2/year}].
\end{align}
The reconstructed stochastic phase equations reproduce the distribution of $\psi_{3\,1}$ in the observed data reasonably well, including the sharp peak near $\psi_{3\,1}/(2\pi)\simeq0.75$.

\par
For $m:n=3:1$, results for the amplitude thresholds $\tilde{R}_{10\,\mathrm{th}}, \tilde{R}_{20\,\mathrm{th}}, \ldots, \tilde{R}_{50\,\mathrm{th}}$ appear in Appendix~\ref{appendix_E}.
The results remained qualitatively similar across these thresholds.

\par
We also reconstructed equations for the frequency ratios $m:n=2:1$, $4:1$, $5:1$, and $5:2$ (see Appendix~\ref{appendix_F}).
In the averaged form of the phase equations for $m:n=2:1$, the phase coupling function $\Gamma_{\mathrm{e}}(\psi_{2\,1})$ is also nonzero, and the velocity of $\psi_{2\,1}$ therefore depends on the phase difference, as in the $3:1$ case.
However, the histogram exhibits no clear preference for any particular value of $\psi_{2\,1}$. 
This is because the phase-dependent modulation in Eq.~(\ref{eq:phase_difference_dynamics_withnoise}) is small relative to the constant term $\omega_{\mathrm a}-2\omega_{\mathrm e}$.
For $4:1$, $5:1$, and $5:2$, the phase coupling function $\Gamma_{\mathrm{e}}(\psi_{mn})$ is nearly zero, and $\psi_{mn}$ evolves at a nearly constant rate.

\section{Discussion}
\label{sec:discussion}

\par
In this study, we quantified the phase dynamics of the annual and ENSO cycles by reconstructing phase equations through Bayesian inference.
The datasets used for Bayesian inference were constructed from phase time series obtained from Koopman eigenfunctions, which were computed using the data-driven Koopman operator approach of \citet{froyland_spectral_2021}.
Using the phase equations, we investigated the seasonal dependence of the duration of ENSO events and the synchronization properties between the two cycles.

\par
In Fig.~\ref{fig:fig2}(a), $Q_{\mathrm e}(\theta_{\mathrm a},\theta_{\mathrm e})$ exhibits distinct bands that slope upward to the right and repeat three times in the $\theta_{\mathrm e}$ direction.
The slope of these bands approximately matches that of the integration paths used for averaging in Eqs.~(\ref{eq:overline_Qa}) and (\ref{eq:overline_Qe}) with $m:n=3:1$.
Thus, averaging along these paths for $m:n=3:1$ is expected to yield a clear resonant component in the coupling function $\Gamma_{\mathrm e}(\psi_{3\,1})$.
Indeed, the equation for the time evolution of $\psi_{3\,1}$ contains nonzero resonant terms that depend on $\psi_{3\,1}$, even though the cutoff condition in Eq.~(\ref{eq:k_cutoff}) limits these terms to the first harmonic (Fig.~\ref{fig:fig3}).
Direct reconstruction of the phase equations in averaged form, which allows higher harmonics to be included, also yielded a qualitatively similar dependence on the phase difference (Fig.~\ref{fig:fig4}(a)).

\par 
Although neither equation shown in Figs.~\ref{fig:fig3} and \ref{fig:fig4}(a) has a stable fixed point corresponding to complete phase locking, the pronounced contrast between regions where $\psi_{3\,1}$ evolves slowly and those where it evolves rapidly suggests statistical $3:1$ phase locking.
The nonuniformity of the observed distribution of the phase difference also supports statistical $3:1$ phase locking, and this distribution was reproduced by the directly reconstructed stochastic phase equations (Fig.~\ref{fig:fig4}(b)).

\par
We also examined synchronization at frequency ratios $m:n=2:1$, $4:1$, $5:1$, and $5:2$, with results presented in Appendix~\ref{appendix_F}.
We found clear statistical phase locking only for $3:1$.
For $2:1$, the phase-dependent modulation of the velocity of $\psi_{2\,1}$ was nonzero but small compared with the constant term, and no clear statistical phase locking was found.
For $4:1$, $5:1$, and $5:2$, the phase-dependent modulation was nearly zero, suggesting that statistical phase locking does not occur.
In particular, the small velocity of $\psi_{4\,1}$ merely reflects the small frequency difference $\omega_{\mathrm a}-4\omega_{\mathrm e}$, rather than phase-dependent coupling.

\par
We reconstructed phase equations from phase time series derived from $g_{\mathrm e}(\omega_\ell)$, which represents the fundamental ENSO mode with a period of approximately four years, and found statistical $3:1$ phase locking with the annual cycle.
In contrast, previous studies defined the ENSO phase from the complex principal component of an ENSO mode obtained by complex empirical orthogonal function analysis \citep{stein_phase_synchronization_2011} or from the Ni\~{n}o 3.4 index \citep{stein_enso_seasonal_synchronization_theory_2014}; both reported statistical $2:1$ phase locking.
In our replication of these methods, the mean ENSO phase frequencies corresponded to periods of approximately $2.47$ and $1.94$ years, respectively.
In these cases, the mean annual--ENSO frequency ratio was close to $2:1$, making it easier for annual forcing to induce (statistical) $2:1$ phase locking.
This is because, when $\omega_{\mathrm a}-2\omega_{\mathrm e}$ is close to zero in Eq.~(\ref{eq:phase_difference_dynamics_withoutnoise}), $\dot{\psi}_{2\,1}$ can approach zero at certain phase differences even when the coupling function $\Gamma_{\mathrm e}(\psi_{2\,1})$ has a small amplitude.
This can produce a pronounced contrast between regions where $\psi_{2\,1}$ evolves rapidly and those where it evolves slowly.

\par
The difference between statistical $2:1$ phase locking in previous studies and statistical $3:1$ phase locking in the present study arises from our use of the pure ENSO mode, which is separated from other modes such as annual--ENSO combination modes.
This separation yields an ENSO phase whose mean frequency differs from those of the ENSO phases used in previous studies; the corresponding mean period is approximately four years in the present study, compared with approximately two years in our replication of the previous studies, as noted in the previous paragraph.
Consequently, the synchronization properties obtained in the present study differ from those reported in previous studies.
For reference, Fig.~\ref{fig:fig5} compares the Ni\~{n}o 3.4 index with its reconstruction using $g_{\mathrm e}(\omega_\ell)$ (see Appendix~\ref{appendix_G} for the reconstruction procedure).
The green shaded regions in the figure highlight examples of double peaks in the original index, to which combination modes contribute \citep{froyland_spectral_2021}.
These double peaks appear as single peaks in the reconstruction, which retains only the fundamental ENSO mode and excludes contributions from combination modes.

\begin{figure}[t]
  \centering
  \includegraphics[width=19pc,angle=0]{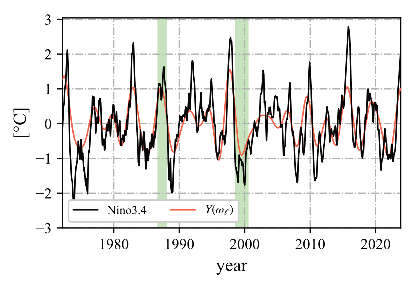}
    \caption{
        Time series obtained from the Koopman eigenfunctions representing the fundamental 4-year ENSO mode.
        The red curve shows $Y(\omega_\ell)$ computed by projecting the Ni\~{n}o 3.4 index onto $g_{\mathrm e}(\omega_\ell)$.
        For details, see Eqs.~(\ref{eq:nino34_projection}) and (\ref{eq:nino34_lag_set}).
        The original Ni\~{n}o 3.4 index is shown in black.
        The green shaded regions (September 1986--January 1988 and July 1998--July 2000) highlight two examples where a double peak in the Ni\~{n}o 3.4 index is represented as a single peak in $Y(\omega_\ell)$.
    }
  \label{fig:fig5}
\end{figure}

\par
Figures~\ref{fig:fig2}(c) and \ref{fig:fig2}(d) show that the mean durations of the El Ni\~{n}o and La Ni\~{n}a stages tend to be longer when the corresponding phase intervals are entered during boreal summer.
In contrast, the mean durations tend to be shorter when the corresponding phase intervals are entered during boreal winter for El Ni\~{n}o and during boreal winter to spring for La Ni\~{n}a.
The results for El Ni\~{n}o are consistent with the conventional picture of El Ni\~{n}o developing during boreal summer and autumn and peaking during boreal winter \citep{rasmusson_variations_tropical_sea_surface_temperature_1982,zebiak_model_el_nino_southern_oscillation_1987,stein_phase_synchronization_2011}, although the present analysis focuses on seasonal modulation of the phase velocity along the ENSO cycle rather than on the intensity of ENSO events.

\par
\begin{figure*}[t]
  \noindent\includegraphics[width=\textwidth,angle=0]{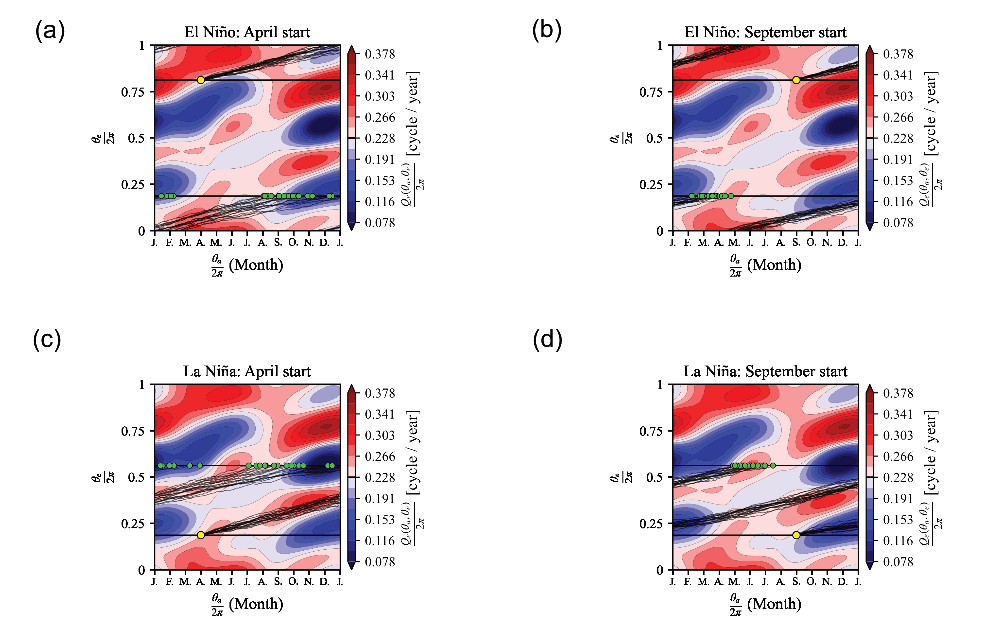}\\
  \caption {
        Illustration of the variability in the durations of the El Ni\~{n}o and La Ni\~{n}a stages, evaluated using Langevin simulations.
        The figure shows trajectories from Langevin simulations starting at the beginning of (a)~the El Ni\~{n}o stages in April; (b)~the El Ni\~{n}o stages in September; (c)~the La Ni\~{n}a stages in April; and (d)~the La Ni\~{n}a stages in September.
        Black curves show representative trajectories overlaid on $Q_{\mathrm e}(\theta_{\mathrm a},\theta_{\mathrm e})$.
        Yellow points mark the initial conditions $(\theta_{\mathrm a}(0)/(2\pi),\theta_{\mathrm e}(0)/(2\pi))$: (a)~$(3/12,13/16)$, (b)~$(8/12,13/16)$, (c)~$(3/12,3/16)$, and (d)~$(8/12,3/16)$.
        Green points mark each trajectory's first arrival at the end of the El Ni\~{n}o or La Ni\~{n}a stages.
        They are plotted at $\theta_{\mathrm e}/(2\pi)=3/16$ in panels (a) and (b) and at $9/16$ in panels (c) and (d).
        For the El Ni\~{n}o stages starting in April, the modulation described by $Q_{\mathrm e}(\theta_{\mathrm a},\theta_{\mathrm e})$ amplifies the spread of trajectories, leading to greater variability in the durations.
        In contrast, for starts in September, this modulation suppresses the spread of trajectories and reduces the variability in the durations.
        The same explanation applies to the La Ni\~{n}a stages.
}
  \label{fig:fig6}
\end{figure*}

Furthermore, Figs.~\ref{fig:fig2}(c) and \ref{fig:fig2}(d) show large variability in the durations of the El Ni\~{n}o and La Ni\~{n}a stages for starts in boreal spring, indicating that uncertainty in duration depends on the starting season.
The bands in $Q_{\mathrm e}(\theta_{\mathrm a},\theta_{\mathrm e})$, particularly pronounced during boreal winter and spring, contribute to this seasonal dependence.
To illustrate this, Fig.~\ref{fig:fig6} shows several trajectories from Langevin simulations overlaid on a color map of $Q_{\mathrm e}(\theta_{\mathrm a},\theta_{\mathrm e})$.
Figures~\ref{fig:fig6}(a) and \ref{fig:fig6}(b) show trajectories during the El Ni\~{n}o stages starting in April and September, respectively, while Figs.~\ref{fig:fig6}(c) and \ref{fig:fig6}(d) show the corresponding trajectories during the La Ni\~{n}a stages.
For both the El Ni\~{n}o and La Ni\~{n}a stages starting in April, the bands amplify the spread of trajectories, leading to large variability in the durations.
In contrast, when these stages start in September, the spread of trajectories is suppressed, resulting in small variability in the durations.

\par
This seasonal dependence of the variability in duration may be related to the spring predictability barrier, characterized by a decline in ENSO predictability during boreal spring~\citep{torrence_annual_cycle_persistence_1998,levine_annual_cycle_enso_growth_rate_2015}.
By combining the phase dynamics with stationary noise, we accounted for the seasonal dependence of the variability in duration without assuming seasonally varying noise intensity.

\par
This study focused on the phase dynamics of the ENSO cycle and did not examine the intensity of El Ni\~{n}o or La Ni\~{n}a events.
\citet{froyland_spectral_2021} noted that larger eigenfunction magnitudes are associated with stronger ENSO cycles.
This suggests that the amplitude used in this study, i.e., $R_{\mathrm e}(t_\ell)$, may provide information about the intensity of ENSO events.
Future work could therefore extend the model to describe the time evolution of both phase and amplitude, drawing on phase-amplitude reduction \citep{wilson_isostable_2016,shirasaka_phase-amplitude_2017,mauroy_global_2018}.
Such an extension could provide a more detailed understanding of the synchronization mechanisms involved in the formation of strong ENSO events.


\acknowledgments
This work was supported by JSPS KAKENHI Grant Numbers JP24K06910, JP25K01160, and JP26K21344.
This work was also supported by Earth Simulator JAMSTEC Proposed Project.
We thank Takeshi Doi, Tsubasa Kohyama, and Shoichiro Kido for useful comments on this work.

%
%
\datastatement
The code and processed data used in this study are available on GitHub at \url{https://github.com/TakahiroArai-33/CODE_ENSO_Phase_Dynamics}.
The ERSSTv5 data are publicly available from NOAA's National Centers for Environmental Information at \url{https://www.ncei.noaa.gov/pub/data/cmb/ersst/v5/}.

\appendix[A]
\setappendixlabel{A}{appendix_A}
\appendixtitle{Overview of the algorithm for computing Koopman eigenfunctions}

\par
This appendix summarizes the computation of the Koopman eigenfunctions used in this study.
We follow the computational procedure of \citet{froyland_spectral_2021} for ENSO analysis and refer to it as the data-driven Koopman operator approach.
This approach constructs a kernel basis from delay-embedded data using nonlinear Laplacian spectral analysis \citep{giannakis_nonlinear_2012} and approximates the Koopman generator in this basis \citep{giannakis_data_driven_2019}.
For the mathematical background of kernel methods for Koopman spectral analysis, see \citet{das_reproducing_2021}.
Similar computational frameworks have been applied to the MJO \citep{lintner_identification_2023} and the QBO \citep{valva_qbo_2025}.

\par
We approximate the Koopman generator from the SST observations without requiring an explicit model of the underlying dynamics.
Snapshots of the SST field are represented by $\boldsymbol{x}(t)=X(\omega(t))$, where $X:\Omega\to\mathbb{R}^d$ is the observation map and $\omega(t)\in\Omega$ is the climate state at time $t$.
The full state $\omega(t)$ is not directly observed.
We denote the state by $\omega_\ell=\omega(t_\ell)$ at the sampling times $t_\ell=\ell\Delta t$, where $\ell=0,1,\ldots,N-1$ and $\Delta t$ is the sampling interval.

\par
Using the sampled SST snapshots $\bm{x}_\ell:=\bm{x}(t_\ell)$, we construct delay-embedded vectors as
\begin{align}
    \label{eq:koopman_delay_embedding}
    \widehat{\bm{x}}_\ell
    &:=
    \left(
        \bm{x}_\ell^{\mathrm T},
        \bm{x}_{\ell-s}^{\mathrm T},
        \ldots,
        \bm{x}_{\ell-(\kappa-1)s}^{\mathrm T}
    \right)^{\mathrm T}
    \in
    \mathbb{R}^{d\kappa}.
\end{align}
Here, $\kappa$ is the number of snapshots in the delay-embedding window and $s$ is the lag spacing in units of the sampling interval $\Delta t$.
In this study, we used $\kappa=48$ and $s=1$, with $d=4868$ and $\Delta t=1/12~[\mathrm{year}]$ for the SST data described in Sec.~\ref{subsec:koopman_results}.

\par
Next, we construct a kernel matrix ${\rm K}\in\mathbb{R}^{N\times N}$ from the delay-embedded data $\{\widehat{\bm{x}}_\ell\}_\ell$:
\begin{align}
    \label{eq:koopman_kernel_matrix}
    {\rm K}_{\ell\ell'}
    &:=
    k
    \left(
        \widehat{\bm{x}}_\ell,
        \widehat{\bm{x}}_{\ell'}
    \right),
\end{align}
where $k(\widehat{\bm{x}}_\ell,\widehat{\bm{x}}_{\ell'})$ is a kernel function measuring similarity between $\widehat{\bm{x}}_\ell$ and $\widehat{\bm{x}}_{\ell'}$.
We then normalize ${\rm K}$ to obtain $\widetilde{{\rm K}}$ and solve the following eigenvalue problem:
\begin{align}
    \label{eq:koopman_kernel_eigenproblem}
    \widetilde{{\rm K}}\bm{\phi}_h
    &=
    \rho_h\bm{\phi}_h.
\end{align}
We retain the $H$ eigenvectors corresponding to the largest eigenvalues $\rho_h$ ($h=0,1,\ldots,H-1$).
We denote the $\ell$th component of $\bm{\phi}_h$ by $\phi_{h,\ell}$, where $\ell=0,1,\ldots,N-1$.
In this study, we retained $H=401$ basis vectors.
Here, $N$ is the number of samples retained after delay embedding and the exclusion of boundary samples for finite differences; for the monthly SST data from January 1970 to December 2025, $N=621$.
For details on the kernel design and normalization, see \citet{froyland_spectral_2021}.

\par
We define the Koopman generator $V$ as the infinitesimal generator of the family of Koopman operators $\{U^t\}$, which is given by
\begin{align}
    \label{eq:koopman_generator_definition_appendix}
    V f
    &:=
    \lim_{t\to0}
    \frac{U^t f-f}{t},
\end{align}
and we define $\widetilde{V}\in\mathbb{R}^{H\times H}$ as the finite-dimensional approximation of $V$ in the basis $\{\bm{\phi}_h\}$.
We add a small diffusion term to $\widetilde{V}$ to improve numerical stability \citep{giannakis_data_driven_2019,froyland_spectral_2021}:
\begin{align}
    \label{eq:koopman_regularized_generator}
    W
    &:=
    \widetilde{V}-\varepsilon\Delta.
\end{align}
Here, $\varepsilon>0$ is a small regularization parameter, and $\Delta\in\mathbb{R}^{H\times H}$ is a diagonal diffusion matrix.
For details of the construction of $\widetilde{V}$ and $W$, see \citet{froyland_spectral_2021}.

\par
Once $W$ has been constructed, we solve the eigenvalue problem
\begin{align}
    \label{eq:koopman_generator_eigenproblem}
    W\bm{u}_j
    &=
    \nu_j \bm{u}_j,
\end{align}
where $(\bm{u}_j,\nu_j)$ is the $j$th eigenpair of $W$.
The eigenvector $\bm{u}_j=(u_{0,j},\ldots,u_{H-1,j})^{\mathrm T}$ contains the expansion coefficients of the $j$th approximate Koopman eigenfunction in the kernel basis $\{\bm{\phi}_h\}_{h=0}^{H-1}$, i.e.,
\begin{align}
    \label{eq:koopman_eigenfunction_timeseries_vector}
    \widehat{\bm{g}}_j
    &:=
    \sum_{h=0}^{H-1}
    u_{h,j}\bm{\phi}_h.
\end{align}
We obtain the following approximation to the Koopman eigenfunction $g_j$ at the sampled state $\omega_\ell$:
\begin{align}
    \label{eq:koopman_eigenfunction_value}
    g_j(\omega_\ell)
    &\simeq
    \widehat{g}_{j,\ell}
    =
    \sum_{h=0}^{H-1}
    u_{h,j}\phi_{h,\ell},
\end{align}
where $\phi_{h,\ell}$ and $\widehat{g}_{j,\ell}$ are the $\ell$th components of $\bm{\phi}_h$ and $\widehat{\bm{g}}_j$, respectively.

\appendix[B]
\setappendixlabel{B}{appendix_B}
\appendixtitle{Bayesian inference of the phase equations}

\par
This appendix describes Bayesian inference of the phase equations in Eqs.~(\ref{eq:phase_equation_annual}) and~(\ref{eq:phase_equation_enso}) from phase time series \citep{stankovski_inference_2012,duggento_dynamical_2012}.

\subsection{Data used for inference}
\label{subsec:appendix_b_data}

\par
For each $\ell$, we define the amplitude at the midpoint as
\begin{align}
    \label{eq:amplitude_midpoint}
    R_{\mathrm{e},\ell}^{\ast}
    &:=
    \frac{
        R_{\mathrm{e}}(t_{\ell+1})
        +
        R_{\mathrm{e}}(t_\ell)
    }{2},
\end{align}
where $R_{\mathrm e}(t_\ell)=|g_{\mathrm e}(\omega_\ell)|$.
We define the set of indices as
\begin{align}
    \label{eq:amplitude_filtered_midpoint_indices}
    D_{\mathrm{amp}}^{\ast}
    &:=
    \left\{
        \ell
        \;\middle|\;
        R_{\mathrm{e},\ell}^{\ast}
        \geq
        \widetilde{R}_{\mathrm{e}}
    \right\},
\end{align}
where $\widetilde{R}_{\mathrm e}$ is the amplitude threshold.
Selection of phase data using this threshold serves as a filter that excludes data with small values of $R_{\mathrm e}$.
The dataset used to reconstruct the phase equations is defined as
\begin{align}
    \label{eq:data_D_centered_difference}
    \mathcal{D}
    &:=
    \left\{
        \left(
            \theta_{\mathrm{a},\ell}^{\ast}\,,
            \theta_{\mathrm{e},\ell}^{\ast}\,,
            \chi_{\mathrm{a},\ell}\,,
            \chi_{\mathrm{e},\ell}
        \right)
    \right\}_{\ell\in D_{\mathrm{amp}}^{\ast}},
\end{align}
where
\begin{align}
    \label{eq:theta_a_ast}
    \theta_{\mathrm{a},\ell}^\ast := \frac{\theta_{\mathrm{a}}(t_\ell + \Delta t) + \theta_{\mathrm{a}}(t_\ell)}{2}, \\
    \label{eq:theta_e_ast}
    \theta_{\mathrm{e},\ell}^\ast := \frac{\theta_{\mathrm{e}}(t_\ell + \Delta t) + \theta_{\mathrm{e}}(t_\ell)}{2}, \\
    \label{eq:dot_theta_a}
    \chi_{\mathrm{a},\ell} := \frac{\theta_{\mathrm{a}}(t_\ell + \Delta t) - \theta_{\mathrm{a}}(t_\ell)}{\Delta t}, \\
    \label{eq:dot_theta_e}
    \chi_{\mathrm{e},\ell} := \frac{\theta_{\mathrm{e}}(t_\ell + \Delta t) - \theta_{\mathrm{e}}(t_\ell)}{\Delta t}.
\end{align}
We define the number of samples retained after filtering as
\begin{align}
    N_{\mathrm s}
    &:=
    \left|D_{\mathrm{amp}}^{\ast}\right|.
\end{align}

\subsection{Statistical model}
\label{subsec:appendix_b_model}

\par
We use the following vector and matrix notation:
\begin{align}
    \label{eq:dot_chi}
    \boldsymbol{\chi}_{\ell}
    := \begin{pmatrix}
        \chi_{\mathrm{a},\ell} \\
        \chi_{\mathrm{e},\ell}
    \end{pmatrix}.
\end{align}
The parameter vector is defined as
\begin{align}
    & \boldsymbol{c}
    :=
    \begin{pmatrix}
        \boldsymbol{c}_{\mathrm{a}} \\
        \boldsymbol{c}_{\mathrm{e}}
    \end{pmatrix}
    \in \mathbb{C}^{2R},
\end{align}
where
$\boldsymbol{c}_{\mathrm{a}}
:= \left(c_{\boldsymbol{m}}^{\mathrm{a}}\right)_{\boldsymbol{m}\in\mathbb{M}\cup\{\boldsymbol{0}\}}$
and
$\boldsymbol{c}_{\mathrm{e}}
:= \left(c_{\boldsymbol{m}}^{\mathrm{e}}\right)_{\boldsymbol{m}\in\mathbb{M}\cup\{\boldsymbol{0}\}}$.
The number of parameters per equation is $R := (2M+1)^2$, which scales as $O(M^2)$.
We further introduce the following matrices:
\begin{align}
    & {\rm H}_{\ell}
    :=
    \begin{pmatrix}
    \boldsymbol{h}_{\ell}^\mathrm{T} & 0 \\
    0 & \boldsymbol{h}_{\ell}^\mathrm{T}
    \end{pmatrix} \in \mathbb{C}^{2 \times 2R},
    \\
    &\partial {\rm H}_{\ell}
    :=
    \begin{pmatrix}
    \frac{\partial}{\partial \theta_{\mathrm{a}}} \boldsymbol{h}_{\ell}^\mathrm{T} & 0 \\
    0 & \frac{\partial}{\partial \theta_{\mathrm{e}}} \boldsymbol{h}_{\ell}^\mathrm{T}
    \end{pmatrix} \in \mathbb{C}^{2 \times 2R},
\end{align}
where $\boldsymbol{h}_{\ell}
:= \left(h_{\boldsymbol{m},\ell}\right)_{\boldsymbol{m}\in\mathbb{M}\cup\{\boldsymbol{0}\}}$
is the vector of Fourier basis functions, with
$h_{\boldsymbol{m},\ell}
:= h_{\boldsymbol{m}}(\theta_{\mathrm{a},\ell}^{*}, \theta_{\mathrm{e},\ell}^{*})$.
In particular, $h_{\boldsymbol{0},\ell}=1$.

\par
We consider the problem of reconstructing the phase equations given by Eqs.~(\ref{eq:phase_equation_annual}) and (\ref{eq:phase_equation_enso}).
The parameters to be estimated are denoted by $\mathcal{M} := \{\boldsymbol{c}, {\rm E}\}$.
Under Gaussian white noise and the midpoint approximation, the likelihood is given by the product of the one-step transition probabilities over $\ell\in D_{\mathrm{amp}}^{\ast}$.
Following \citet{duggento_inferential_2008,stankovski_inference_2012,duggento_dynamical_2012}, the negative log-likelihood is given by
\begin{align}
    \label{eq:negative_log_likelihood}
    S \left( \mathcal{D} \mid \mathcal{M} \right)
    &= \frac{N_{\mathrm s}}{2} \ln |{\rm E}|
    + \frac{\Delta t}{2} \sum_{\ell\in D_{\mathrm{amp}}^{\ast}}
    \Biggl[
    \boldsymbol{1}^{\mathrm T}
    \left(\partial {\rm H}_{\ell} \right)
    \boldsymbol{c}
    \nonumber\\
    &\qquad
    +
    \left(
    \boldsymbol{\chi}_{\ell}
    -
    {\rm H}_{\ell}\, \boldsymbol{c}
    \right)^{\dagger}
    {\rm E}^{-1}
    \left(
    \boldsymbol{\chi}_{\ell}
    -
    {\rm H}_{\ell}\, \boldsymbol{c}
    \right)
    \Biggr], 
\end{align}
where $\boldsymbol{1} = (1,1)^\mathrm{T}$.
The term $\boldsymbol{1}^{\mathrm T}(\partial {\rm H}_{\ell})\boldsymbol{c}$ arises from the Jacobian of the transformation from the noise to the phase increments, with the Jacobian determinant approximated by the product of the diagonal entries of the Jacobian matrix \citep{luchinsky_inferential_2008}.

\par
We estimate the parameters $\mathcal{M}$ from the dataset $\mathcal{D}$.
The prior distribution for $\boldsymbol{c}$ is assumed to be Gaussian.
\begin{align}
    \label{eq:prior_distribution}
    \boldsymbol{c} \sim \mathcal{N}(\boldsymbol{c}_{\mathrm{prior}}, \Sigma_{\mathrm{prior}}).
\end{align}
The posterior distribution is determined in accordance with Bayes' theorem:
\begin{align}
    \label{eq:bayes}
    P_{\mathrm{post}} \left( \mathcal{M} \mid \mathcal{D} \right)
    \propto 
    L \left( \mathcal{D} \mid \mathcal{M} \right)
    P_{\mathrm{prior}} \left( \mathcal{M} \right),
\end{align}
where $L \left( \mathcal{D} \mid \mathcal{M} \right) := \exp \left[ - S \left( \mathcal{D} \mid \mathcal{M} \right) \right]$ is the likelihood.
For fixed ${\rm E}$, the Gaussian prior distribution for $\boldsymbol{c}$ is conjugate to the likelihood; therefore, the conditional posterior of $\boldsymbol{c}$ is also Gaussian.
Equation~(\ref{eq:bayes}) is expressed in terms of $\mathcal{M}$ for notational convenience, although we specify no prior distribution for ${\rm E}$.

\subsection{Bayesian update rules}
\label{subsec:appendix_b_estimation}

\par
The parameter vector $\boldsymbol{c}$ is determined by maximizing the posterior distribution, whereas ${\rm E}$ is determined by minimizing the negative log-likelihood.
The model parameters are thus determined by iterating the following update rules:
\begin{align}
    \label{eq:update_E}
    & {\rm E}
    = \frac{\Delta t}{N_{\mathrm s}} \sum_{\ell\in D_{\mathrm{amp}}^{\ast}}
    \bigl( 
        \boldsymbol{\chi}_{\ell} - {\rm H}_{\ell}\, \boldsymbol{c}
    \bigr)
    \bigl( 
        \boldsymbol{\chi}_{\ell} - {\rm H}_{\ell}\, \boldsymbol{c}
    \bigr)^\dagger, \\
    \label{eq:update_Xi}
    & \Xi
    = \Xi_{\mathrm{prior}}
    + \Delta t \sum_{\ell\in D_{\mathrm{amp}}^{\ast}} {\rm H}_{\ell}^\dagger {\rm E}^{-1} {\rm H}_{\ell}, \\
    \label{eq:update_r}
    & \boldsymbol{r}
    = \Xi_{\mathrm{prior}}\, \boldsymbol{c}_{\mathrm{prior}}
    + \Delta t \sum_{\ell\in D_{\mathrm{amp}}^{\ast}} {\rm H}_{\ell}^\dagger {\rm E}^{-1} \boldsymbol{\chi}_{\ell} \nonumber \\
    & \quad \quad - \frac{\Delta t}{2} \sum_{\ell\in D_{\mathrm{amp}}^{\ast}} (\partial {\rm H}_{\ell})^{\dagger} \boldsymbol{1}, \\
    \label{eq:update_c}
    & \boldsymbol{c}
    = \Xi^{-1}
    \boldsymbol{r},
\end{align}
where $\Xi = \Sigma^{-1}$ and $\Xi_{\mathrm{prior}} = \Sigma_{\mathrm{prior}}^{-1}$.
Starting from $\boldsymbol{c}=\boldsymbol{c}_{\mathrm{prior}}$ and $\Sigma = \Sigma_{\mathrm{prior}}$, we iteratively evaluate Eqs.~(\ref{eq:update_E})--(\ref{eq:update_c}) until we obtain the stationary point of $S \left(\mathcal{D} \mid \mathcal{M} \right)$.

\par 
The hyperparameters were set as follows:
\begin{align}
    \boldsymbol{c}_{\mathrm{prior}} &= \boldsymbol{0}, \\
    \Sigma_{\mathrm{prior}} &= 1.0\times10^{6}\,{\rm I}.
\end{align}

\appendix[C]
\setappendixlabel{C}{appendix_C}
\appendixtitle{Bayesian inference for the averaged form of phase equations}

\par
This appendix describes Bayesian inference for direct reconstruction of the averaged form of the phase equations (Eqs.~(\ref{eq:averaged_phase_equation_annual}) and (\ref{eq:averaged_phase_equation_enso})) from phase time series~\citep{ota_direct_2014,onojima_dynamical_2018,ota_interaction_2020,arai_extracting_2022,furukawa_bayesian_2025}.

\subsection{Data used for inference}
\label{subsec:appendix_c_data}

\par
We use the index set $D_{\mathrm{amp}}^{\ast}$ defined in Eq.~(\ref{eq:amplitude_filtered_midpoint_indices}) to select phase data for inference.
Therefore, we apply the same filter with threshold $\widetilde{R}_{\mathrm e}$.
The dataset used to reconstruct the phase equations is defined as
\begin{align}
    \label{eq:data_D_forward_difference}
    \mathcal{D}_{\mathrm{p}}
    :=
    \left\{ 
        \left(
            \psi_\ell,
            \chi_{\mathrm{p},\ell}
        \right)
    \right\}_{\ell\in D_{\mathrm{amp}}^{\ast}},
\end{align}
where $\mathrm{p}\in\{\mathrm{a},\mathrm{e}\}$.
The phase difference $\psi_\ell$ and phase velocities $\chi_{\mathrm{p},\ell}$ are defined as
\begin{align}
    \label{eq:forward_psi}
    \psi_\ell
    &:=
    \theta_{\mathrm a,\ell}^{\ast}
    -\frac{m}{n}\theta_{\mathrm e,\ell}^{\ast},
    \\
    \label{eq:forward_chi_p}
    \chi_{\mathrm{p},\ell}
    &:=
    \frac{\theta_{\mathrm{p}}(t_{\ell+1})-\theta_{\mathrm{p}}(t_\ell)}{\Delta t}.
\end{align}
The number of samples retained after filtering is $N_{\mathrm s}=\left|D_{\mathrm{amp}}^{\ast}\right|$.
The phase difference $\psi_\ell$ depends on the frequency ratio $m:n$, but we omit $m$ and $n$ from the notation.

\subsection{Statistical model}
\label{subsec:appendix_c_model}

\par
From the noise covariance matrix ${\rm E}$ obtained by reconstructing the phase equations, Eqs.~(\ref{eq:phase_equation_annual}) and (\ref{eq:phase_equation_enso}), we found that the correlation between the noise terms in the two equations is weak (Eq.~(\ref{eq:estimated_noise_intensity_matrix}) in Sec.~\ref{subsec:phase_dynamics_results}).
On the basis of this result, we assume that the noise terms in Eqs.~(\ref{eq:averaged_phase_equation_annual}) and (\ref{eq:averaged_phase_equation_enso}) are independent.

\par
We define the coefficient vector and the basis vector as
\begin{align}
    \label{eq:bayes2_parameter_vector}
    \boldsymbol{w}_{\mathrm{p}}
    &:=
    \begin{aligned}[t]
        &\bigl(
            \omega_{\mathrm{p}},\;
            a_1^{\mathrm{p}},\;
            b_1^{\mathrm{p}},\;
            a_2^{\mathrm{p}},\;
            b_2^{\mathrm{p}},
        \\
        &\quad
            \ldots,\;
            a_M^{\mathrm{p}},\;
            b_M^{\mathrm{p}}
        \bigr)^{\mathrm T}
        \in \mathbb{R}^{2M+1},
    \end{aligned}
    \\
    \label{eq:bayes2_feature_vector}
    \boldsymbol{f}_\ell
    &:=
    \begin{aligned}[t]
        &\bigl(
            1,\;
            \cos\psi_\ell,\;
            \sin\psi_\ell,\;
            \cos(2\psi_\ell),\;
            \sin(2\psi_\ell),
        \\
        &\quad
            \ldots,\;
            \cos(M\psi_\ell),\;
            \sin(M\psi_\ell)
        \bigr)^{\mathrm T}
        \in \mathbb{R}^{2M+1}.
    \end{aligned}
\end{align}
The number of coefficients to be estimated per phase equation is $R':=2M+1$, which scales as $O(M)$.

\par
We consider the problem of reconstructing the phase equations given by Eqs.~(\ref{eq:averaged_phase_equation_annual}) and (\ref{eq:averaged_phase_equation_enso}).
We define the residual variance as $v_{\mathrm{p}}:=\sigma_{\mathrm{p}}^2/\Delta t$.
The parameters to be estimated are $\mathcal{M}_{\mathrm{p}}:=\{\boldsymbol{w}_{\mathrm{p}},v_{\mathrm{p}}\}$.
The negative log-likelihood is given by
\begin{align}
    \label{eq:bayes2_negative_log_likelihood}
    S_{\mathrm{p}}
    \left(
        \mathcal{D}_{\mathrm{p}}
        \mid
        \mathcal{M}_{\mathrm{p}}
    \right)
    &:=
    -\log L
    \left(
        \mathcal{D}_{\mathrm{p}}
        \mid
        \mathcal{M}_{\mathrm{p}}
    \right)
    \nonumber\\
    &=
    \frac{N_{\mathrm s}}{2}\log(2\pi v_{\mathrm{p}})
    + \frac{1}{2v_{\mathrm{p}}}
    \sum_{\ell\in D_{\mathrm{amp}}^{\ast}}
    \left(
        \chi_{\mathrm{p},\ell}
        -
        \boldsymbol{f}_\ell^{\mathrm T}
        \boldsymbol{w}_{\mathrm{p}}
    \right)^2 .
\end{align}

\par
We use a Gaussian-inverse-gamma prior distribution for the coefficients and residual variance.
Whereas we obtained a point estimate of the noise covariance when reconstructing Eqs.~(\ref{eq:phase_equation_annual}) and (\ref{eq:phase_equation_enso}), here we also account for uncertainty in the residual variance through its posterior distribution.
\begin{align}
    \label{eq:bayes2_prior}
    P_{\mathrm{prior}}
    \left(
        \mathcal{M}_{\mathrm{p}}
    \right)
    &=
    \begin{aligned}[t]
    &\mathcal{N}
    \left(
        \boldsymbol{w}_{\mathrm{p}}
        \;\middle|\;
        \boldsymbol{w}_{\mathrm{p}}^{\mathrm{prior}},
        v_{\mathrm{p}}\Sigma_{\mathrm{p}}^{\mathrm{prior}}
    \right)
    \\
    &\quad\times\mathcal{IG}
    \left(
        v_{\mathrm{p}}
        \;\middle|\;
        \alpha_{\mathrm{p}}^{\mathrm{prior}},
        \beta_{\mathrm{p}}^{\mathrm{prior}}
    \right),
    \end{aligned}
    \\
    \label{eq:bayes2_inverse_gamma}
    \mathcal{IG}(x\mid \alpha,\beta)
    &=
    \frac{\beta^\alpha}{\Gamma(\alpha)}
    x^{-\alpha-1}
    \exp\left(-\frac{\beta}{x}\right).
\end{align}
By Bayes' theorem, the posterior distribution is given by
\begin{align}
    \label{eq:bayes2_posterior}
    P_{\mathrm{post}}
    \left(
        \mathcal{M}_{\mathrm{p}}
        \mid
        \mathcal{D}_{\mathrm{p}}
    \right)
    &\propto
    L\left(\mathcal{D}_{\mathrm{p}}\mid\mathcal{M}_{\mathrm{p}}\right)
    P_{\mathrm{prior}}\left(\mathcal{M}_{\mathrm{p}}\right).
\end{align}
Because the Gaussian-inverse-gamma prior distribution is conjugate to the likelihood, the posterior also has a Gaussian-inverse-gamma distribution.
Parameters are estimated independently for each $\mathrm{p}\in\{\mathrm{a},\mathrm{e}\}$.

\subsection{Bayesian update rules}
\label{subsec:appendix_c_estimation}

\par
The coefficient vector $\boldsymbol{w}_{\mathrm p}$ is determined as the mean of the Gaussian posterior distribution conditional on $v_{\mathrm p}$ and is therefore given by
\begin{align}
    \label{eq:bayes2_update_sigma}
    \Sigma_{\mathrm{p}}
    &=
    \left[
        \left(\Sigma_{\mathrm{p}}^{\mathrm{prior}}\right)^{-1}
        + \sum_{\ell\in D_{\mathrm{amp}}^{\ast}}
        \boldsymbol{f}_\ell
        \boldsymbol{f}_\ell^{\mathrm T}
    \right]^{-1},
    \\
    \label{eq:bayes2_update_w}
    \boldsymbol{w}_{\mathrm{p}}
    &=
    \Sigma_{\mathrm{p}}
    \left[
        \sum_{\ell\in D_{\mathrm{amp}}^{\ast}}
        \boldsymbol{f}_\ell
        \chi_{\mathrm{p},\ell}
        + \left(\Sigma_{\mathrm{p}}^{\mathrm{prior}}\right)^{-1}
        \boldsymbol{w}_{\mathrm{p}}^{\mathrm{prior}}
    \right].
\end{align}
The hyperparameters of the distribution of $v_{\mathrm p}$ are updated as follows:
\begin{align}
    \label{eq:bayes2_update_alpha}
    \alpha_{\mathrm{p}}
    &=
    \alpha_{\mathrm{p}}^{\mathrm{prior}}
    + \frac{N_{\mathrm s}}{2}. 
    \\
    \label{eq:bayes2_update_beta}
    \beta_{\mathrm{p}}
    &=
    \beta_{\mathrm{p}}^{\mathrm{prior}}
    + \frac{1}{2}
    \left[
        \sum_{\ell\in D_{\mathrm{amp}}^{\ast}}
        \chi_{\mathrm{p},\ell}^2
        + \left(\boldsymbol{w}_{\mathrm{p}}^{\mathrm{prior}}\right)^{\mathrm T}
        \left(\Sigma_{\mathrm{p}}^{\mathrm{prior}}\right)^{-1}
        \boldsymbol{w}_{\mathrm{p}}^{\mathrm{prior}}
        \right.
        \nonumber\\
        &\qquad\qquad\left.
        - \boldsymbol{w}_{\mathrm{p}}^{\mathrm T}
        \Sigma_{\mathrm{p}}^{-1}
        \boldsymbol{w}_{\mathrm{p}}
    \right].
\end{align}
The residual variance $v_{\mathrm p}$ is determined as the mean of its posterior distribution, $v_{\mathrm p}=\beta_{\mathrm p}/(\alpha_{\mathrm p}-1)$.
Accordingly, the noise intensity in the averaged form of the phase equations is determined by
\begin{align}
    \sigma_{\mathrm p}^2
    &= \Delta t\,v_{\mathrm p}
    = \frac{\Delta t\,\beta_{\mathrm p}}{\alpha_{\mathrm p}-1}.
\end{align}

\par 
The hyperparameters were set as follows:
\begin{align}
    \label{eq:bayes2_prior_initial_w}
    \boldsymbol{w}_{\mathrm{p}}^{\mathrm{prior}}
    &:=
    \left(
        \left\langle \chi_{\mathrm{p},\ell}\right\rangle_{\ell},
        0,
        \ldots,
        0
    \right)^{\mathrm T}
    \in \mathbb{R}^{2M+1},
    \\
    \label{eq:bayes2_prior_initial_covariance}
    \Sigma_{\mathrm{p}}^{\mathrm{prior}}
    &:=
    10^{-\lambda_{\mathrm{p}}}{\rm I},
    \\
    \label{eq:bayes2_prior_initial_alpha_beta}
    \alpha_{\mathrm{p}}^{\mathrm{prior}}
    &:=
    2,
    \qquad
    \beta_{\mathrm{p}}^{\mathrm{prior}}
    :=
    2,
\end{align}
where $\langle\cdot\rangle_\ell$ denotes the average over $\ell\in D_{\mathrm{amp}}^{\ast}$.

\subsection{Model selection}
\label{subsec:appendix_c_model_selection}

\par
The Gaussian-inverse-gamma prior allows the model evidence to be evaluated analytically by integrating over the coefficients and residual variance \citep{bishop_pattern_2006}.
We use model selection based on the model evidence to avoid overfitting fluctuations arising from noise \citep{mackay_bayesian_1992}.
Specifically, we select $\lambda_{\mathrm p}$, which controls the scale of the prior covariance of the coefficients in Eq.~(\ref{eq:bayes2_prior_initial_covariance}).
Selecting $\lambda_{\mathrm p}$ based on the observed data adjusts how strongly the prior influences the estimation of $\boldsymbol{w}_{\mathrm p}$.

\par
We define the candidate set as $\mathcal{L}:=\{0,1,2,\ldots,8\}$.
For each candidate $\lambda'\in\mathcal{L}$, we estimate the parameters using Eqs.~(\ref{eq:bayes2_update_sigma})--(\ref{eq:bayes2_update_beta}).
Thus, we obtain phase equations (Eqs.~(\ref{eq:averaged_phase_equation_annual}) and (\ref{eq:averaged_phase_equation_enso})) corresponding to each $\lambda'\in\mathcal{L}$.
To compare phase equations reconstructed for different $\lambda'\in\mathcal{L}$, we calculate their model evidence as follows:
\begin{align}
    \label{eq:bayes2_model_evidence}
    P_{\mathrm{ME}}^{\mathrm{p}}
    \left(
        \mathcal{D}_{\mathrm{p}} \mid \lambda'
    \right)
    &=
    \frac{
        L\left(\mathcal{D}_{\mathrm{p}}\mid\mathcal{M}_{\mathrm{p}}\right)
        P_{\mathrm{prior}}
        \left(
            \mathcal{M}_{\mathrm{p}}
            \mid \lambda'
        \right)
    }{
        P_{\mathrm{post}}
        \left(
            \mathcal{M}_{\mathrm{p}}
            \mid \mathcal{D}_{\mathrm{p}},\lambda'
        \right)
    }.
\end{align}
We select the hyperparameter that maximizes the model evidence as
\begin{align}
    \label{eq:bayes2_lambda_selection}
    \lambda_{\mathrm{p}}
    &=
    \operatorname*{arg\,max}_{\lambda'\in\mathcal{L}}
    P_{\mathrm{ME}}^{\mathrm{p}}
    \left(\mathcal{D}_{\mathrm{p}} \mid \lambda'\right).
\end{align}
We then use phase equations reconstructed with the prior covariance scale in Eq.~(\ref{eq:bayes2_prior_initial_covariance}) set by the selected $\lambda_{\mathrm p}$.

\appendix[D]
\setappendixlabel{D}{appendix_D}
\appendixtitle{Phase equations for different amplitude thresholds}

\par
\begin{figure*}[t]
  \noindent\includegraphics[width=0.9\textwidth,angle=0]{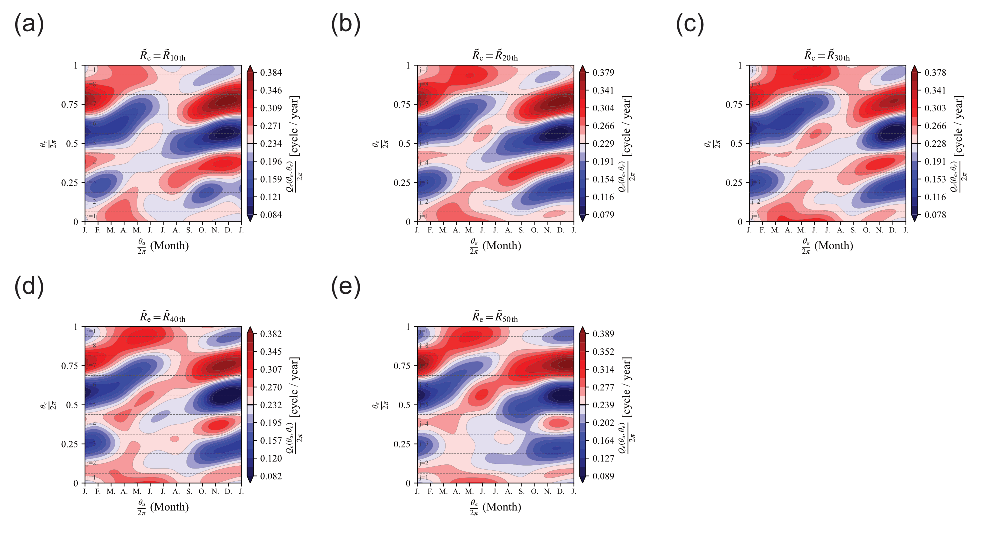}\\
  \caption{
        Reconstructed function $Q_{\mathrm e}(\theta_{\mathrm a},\theta_{\mathrm e})$ for different amplitude thresholds.
        The amplitude threshold $\tilde{R}_{\mathrm e}$ is set to
        (a)~$\tilde{R}_{10\,\mathrm{th}}$,
        (b)~$\tilde{R}_{20\,\mathrm{th}}$,
        (c)~$\tilde{R}_{30\,\mathrm{th}}$,
        (d)~$\tilde{R}_{40\,\mathrm{th}}$, and
        (e)~$\tilde{R}_{50\,\mathrm{th}}$.
        }
    \label{fig:figS1}
\end{figure*}

Figure~\ref{fig:figS1} shows the reconstructed function $Q_{\mathrm e}(\theta_{\mathrm a},\theta_{\mathrm e})$ for amplitude thresholds $\tilde{R}_{\mathrm e}\in\{\tilde{R}_{10\,\mathrm{th}},\tilde{R}_{20\,\mathrm{th}},\ldots,\tilde{R}_{50\,\mathrm{th}}\}$.
The results remain qualitatively similar across these thresholds, indicating robustness to the choice of threshold within this range.
A higher threshold, such as $\tilde{R}_{\mathrm e}=\tilde{R}_{70\,\mathrm{th}}$, leaves some regions of the $(\theta_{\mathrm a},\theta_{\mathrm e})$ plane without data points, which can lead to overfitting.

\appendix[E]
\setappendixlabel{E}{appendix_E}
\appendixtitle{Averaged form of the phase equations for different amplitude thresholds}

\par
\begin{figure}[t]
  \centering
  \includegraphics[width=19pc,angle=0]{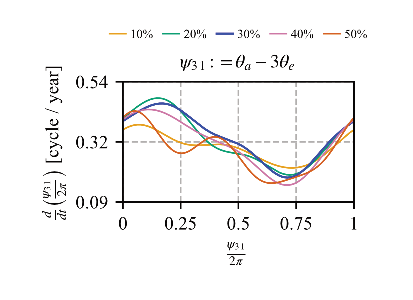}
  \caption{
    Results of direct reconstruction of the averaged form of the phase equations for different amplitude thresholds.
    The orange, green, blue, purple, and red curves correspond to amplitude thresholds of $\tilde{R}_{10\,\mathrm{th}}$, $\tilde{R}_{20\,\mathrm{th}}$, $\tilde{R}_{30\,\mathrm{th}}$, $\tilde{R}_{40\,\mathrm{th}}$, and $\tilde{R}_{50\,\mathrm{th}}$, respectively.
        }
    \label{fig:figS2}
\end{figure}

Figure~\ref{fig:figS2} shows the deterministic part of Eq.~(\ref{eq:phase_difference_dynamics_withnoise}), which describes the time evolution of the phase difference.
The colored curves correspond to the reconstructed equations for different amplitude thresholds $\tilde{R}_{\mathrm e}\in\{\tilde{R}_{10\,\mathrm{th}},\tilde{R}_{20\,\mathrm{th}},\ldots,\tilde{R}_{50\,\mathrm{th}}\}$.
The results remain qualitatively similar across these thresholds, indicating robustness to the choice of threshold within this range.

\appendix[F]
\setappendixlabel{F}{appendix_F}
\appendixtitle{Averaged form of the phase equations for different frequency ratios}

\par
\begin{figure*}[t]
  \noindent\includegraphics[width=\textwidth,angle=0]{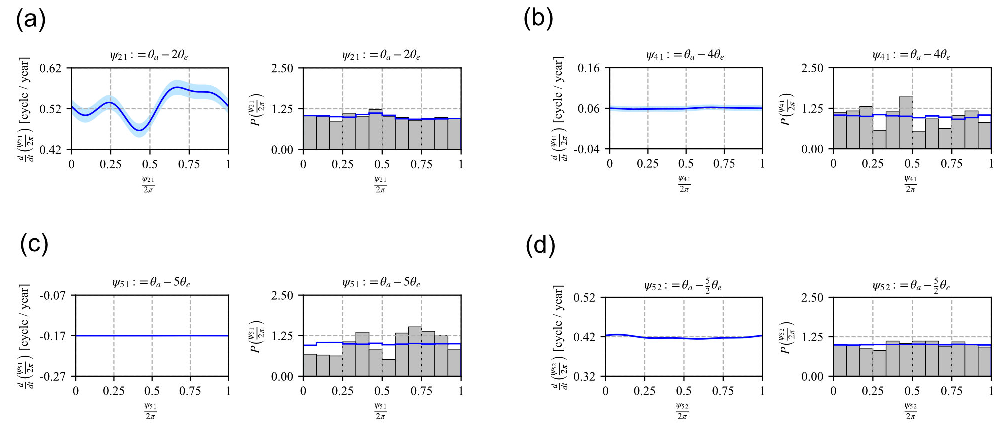}\\
  \caption{
        Results of direct reconstruction of the phase equations in averaged form for different frequency ratios.
        The panels correspond to the frequency ratios $m:n=$
        $2:1$~(a),
        $4:1$~(b),
        $5:1$~(c), and
        $5:2$~(d).
        In each panel, the equation for the phase difference and the histogram are shown in the same format as in Figs.~\ref{fig:fig4}(a) and \ref{fig:fig4}(b), respectively.
        }
    \label{fig:figS3}
\end{figure*}

In Sec.~\ref{subsec:direct_inference_results}, we present the directly reconstructed equation describing the time evolution of the phase difference, Eq.~(\ref{eq:phase_difference_dynamics_withnoise}), only for the frequency ratio $m:n=3:1$.
The equations for other frequency ratios are not plotted there.
Figure~\ref{fig:figS3} shows the deterministic parts of these equations and the corresponding histograms for $m:n=2:1$, $4:1$, $5:1$, and $5:2$.

\par
To assess synchronization at different frequency ratios, we examine the deterministic part of Eq.~(\ref{eq:phase_difference_dynamics_withnoise}), which governs the time evolution of the phase difference $\psi_{mn}$.
Table~\ref{tab:direct_averaging_comparison} lists the constant term $\omega_\psi$ and the amplitude $\Delta\Gamma_{\psi}$ of the phase-dependent coupling term for each frequency ratio.
These quantities are defined as
\begin{align}
    \label{eq:omega_psi}
    \omega_\psi
    &:= \omega_{\mathrm a}-\frac{m}{n}\omega_{\mathrm e},
    \\
    \label{eq:delta_gamma_psi}
    \Delta\Gamma_{\psi}
    &:=
    \max_{\psi_{mn}}\Gamma_{\psi}(\psi_{mn})
    -
    \min_{\psi_{mn}}\Gamma_{\psi}(\psi_{mn}),
\end{align}
where 
\begin{align}
    \Gamma_{\psi}(\psi_{mn})
    &:= \Gamma_{\mathrm a}(\psi_{mn})
    -\frac{m}{n}\Gamma_{\mathrm e}(\psi_{mn}).
\end{align}
If $\Delta\Gamma_\psi<|\omega_\psi|$, the deterministic phase-difference equation has no fixed point, and complete phase locking does not occur \citep{pikovsky_synchronization_2001}.
Statistical phase locking may nevertheless arise from phase-dependent slowing of the phase difference.
The relative magnitudes of $|\omega_\psi|$ and $\Delta\Gamma_\psi$ are therefore important for assessing statistical phase locking.
A small $|\omega_\psi|$ alone does not imply (statistical) phase locking if phase-dependent modulation is negligible.

\begin{table*}[p]
    \caption{
        Values of $\omega_\psi$ and $\Delta\Gamma_\psi$ for $m:n=2:1$, $5:2$, $3:1$, $4:1$, and $5:1$.
        These quantities are defined in Eqs.~(\ref{eq:omega_psi}) and (\ref{eq:delta_gamma_psi}), respectively.
    }
  \label{tab:direct_averaging_comparison}
  \centering
  \begin{tabular}{lrrrrr}
    \hline
    $m:n$ & $2:1$ & $5:2$ & $3:1$ & $4:1$ & $5:1$ \\
    \hline
    $\omega_\psi/(2\pi) \quad \mathrm{[cycle / year]}$
      & $0.5333$ & $0.4142$ & $0.3321$ & $0.0626$ & $-0.1718$ \\
    $\Delta\Gamma_\psi/(2\pi) \quad \mathrm{[cycle / year]}$
      & $0.1060$ & $0.0110$ & $0.2775$ & $0.0046$ & $0.0003$ \\
    \hline
  \end{tabular}
\end{table*}

\par 
Table~\ref{tab:direct_averaging_comparison} shows that the phase-dependent modulation for $m:n=2:1$ is nonzero but small compared with the constant term.
For $m:n=4:1$, $5:1$, and $5:2$, the modulation is nearly zero, and the phase difference evolves at a nearly constant rate.
The phase-dependent modulation is comparable in magnitude to the constant term only for $m:n=3:1$.
The modulation due to $\Gamma_{\mathrm e}(\psi_{3\,1})$ gives rise to statistical phase locking, as described in Sec.~\ref{subsec:direct_inference_results}.

\par 
Figure~\ref{fig:figS3} also shows, together with the equations, histograms of the phase differences obtained from the SST field data and from Langevin simulations of the reconstructed stochastic phase equations.
For $m:n=2:1$, the histograms are nearly flat, although the reconstructed equation contains phase-dependent modulation.
As discussed above, this is because $\Delta\Gamma_\psi$ is small compared with $|\omega_\psi|$.
For $m:n=5:2$, the histograms are also nearly flat, consistent with the nearly constant velocity of the phase difference.
For $m:n=4:1$ and $5:1$, the histograms obtained from the phase time-series data are not flat.
The limited observational record explains this nonuniformity.
Slowly evolving phase differences yield few observed cycles, potentially biasing the histograms.


\appendix[G]
\setappendixlabel{G}{appendix_G}
\appendixtitle{Reconstruction of the fundamental ENSO component in the Ni\~{n}o 3.4 index}

\par
Let $\widehat{\bm{g}}_{\mathrm e}$ denote the vector of approximate values of the Koopman eigenfunction representing the fundamental 4-year ENSO mode (the subscript $\mathrm e$ is used in place of $j$ in Eq.~(\ref{eq:koopman_eigenfunction_timeseries_vector})).
The corresponding eigenvalue $\nu_{\mathrm e}$ satisfies $\operatorname{Im}\nu_{\mathrm e}/(2\pi)\approx0.25~[\mathrm{cycle/year}]$, as specified in Eq.~(\ref{eq:koopman_eigenfrequency_enso}).
The $\ell$th component of $\widehat{\bm{g}}_{\mathrm e}$, denoted by $\widehat{g}_{\mathrm e,\ell}$, is given by Eq.~(\ref{eq:koopman_eigenfunction_value}).
We project a given scalar observable onto this eigenfunction and its complex conjugate at each lag and average the resulting reconstruction over the delay-embedding window to obtain a real-valued observable $Y:\Omega\to\mathbb{R}$:
\begin{align}
    \label{eq:nino34_projection}
    & Y(\omega_\ell)
    :=
    \frac{2}{|\mathcal{W}_\ell|}
    \operatorname{Re}
    \left[
        \sum_{q\in\mathcal{W}_\ell}
        m_{\mathrm e}^{(q)}
        \widehat{g}_{\mathrm e,\ell+q}
    \right], \\
    \label{eq:nino34_lag_set}
    & \mathcal{W}_\ell
    :=
    \left\{
        q\in\mathbb{Z}:
        -\frac{\kappa}{2}\leq q\leq\frac{\kappa}{2}-1
        \quad
        0\leq\ell+q<N
    \right\}.
\end{align}
Here, $m_{\mathrm e}^{(q)}$ is the projection coefficient used in the average over the delay-embedding window, and $q$ denotes the lag.
We applied this procedure to the Ni\~{n}o 3.4 index to extract its fundamental 4-year ENSO component (Fig.~\ref{fig:fig5}).
For details of the projection procedure, see \citet{froyland_spectral_2021}.

\clearpage




\bibliographystyle{ametsocV6}
\bibliography{references}

@article{huang_ersstv5_2017,
  author = {Huang, B. and Thorne, P. W. and Banzon, V. F. and Boyer, T. and Chepurin, G. and Lawrimore, J. H. and Menne, M. J. and Smith, T. M. and Vose, R. S. and Zhang, H.-M.},
  title = {Extended Reconstructed Sea Surface Temperature, Version 5 ({ERSSTv5}): Upgrades, Validations, and Intercomparisons},
  journal = {Journal of Climate},
  year = {2017},
  volume = {30},
  pages = {8179--8205},
  doi = {10.1175/JCLI-D-16-0836.1}
}

@article{rayner_hadisst_2003,
  author = {Rayner, N. A. and Parker, D. E. and Horton, E. B. and Folland, C. K. and Alexander, L. V. and Rowell, D. P. and Kent, E. C. and Kaplan, A.},
  title = {Global analyses of sea surface temperature, sea ice, and night marine air temperature since the late nineteenth century},
  journal = {Journal of Geophysical Research: Atmospheres},
  year = {2003},
  volume = {108},
  number = {D14},
  eid = {4407},
  doi = {10.1029/2002JD002670}
}

@article{hirahara_cobesst2_2014,
  author = {Hirahara, S. and Ishii, M. and Fukuda, Y.},
  title = {Centennial-scale sea surface temperature analysis and its uncertainty},
  journal = {Journal of Climate},
  year = {2014},
  volume = {27},
  pages = {57--75},
  doi = {10.1175/JCLI-D-12-00837.1}
}

@book{kuramoto_chemical_1984,
  author = {Kuramoto, Y.},
  title = {Chemical Oscillations, Waves, and Turbulence},
  year = {1984},
  publisher = {Springer},
  address = {New York},
  doi = {10.1007/978-3-642-69689-3}
}

@book{winfree_geometry_1980,
  author = {Winfree, A. T.},
  title = {The Geometry of Biological Time},
  year = {1980},
  publisher = {Springer},
  address = {New York},
  doi = {10.1007/978-3-662-22492-2}
}

@book{pikovsky_synchronization_2001,
  author = {Pikovsky, A. and Rosenblum, M. and Kurths, J.},
  title = {Synchronization: A Universal Concept in Nonlinear Sciences},
  year = {2001},
  publisher = {Cambridge University Press},
  address = {Cambridge},
  doi = {10.1017/CBO9780511755743}
}

@article{schafer_synchronization_1999,
  author = {Sch{\"a}fer, C. and Rosenblum, M. G. and Abel, H.-H. and Kurths, J.},
  title = {Synchronization in the human cardiorespiratory system},
  journal = {Physical Review E},
  year = {1999},
  volume = {60},
  pages = {857--870},
  doi = {10.1103/PhysRevE.60.857}
}

@article{nakao_phase_2016,
  author = {Nakao, H.},
  title = {Phase reduction approach to synchronisation of nonlinear oscillators},
  journal = {Contemporary Physics},
  year = {2016},
  volume = {57},
  pages = {188--214},
  doi = {10.1080/00107514.2015.1094987}
}

@article{pietras_network_2019,
  author = {Pietras, B. and Daffertshofer, A.},
  title = {Network dynamics of coupled oscillators and phase reduction techniques},
  journal = {Physics Reports},
  year = {2019},
  volume = {819},
  pages = {1--105},
  doi = {10.1016/j.physrep.2019.06.001}
}

@article{guckenheimer_isochrons_1975,
  author = {Guckenheimer, J.},
  title = {Isochrons and phaseless sets},
  journal = {Journal of Mathematical Biology},
  year = {1975},
  volume = {1},
  pages = {259--273},
  doi = {10.1007/BF01273747}
}

@article{mauroy_extreme_2015,
  author = {Mauroy, A. and Mezi{\'c}, I.},
  title = {Extreme phase sensitivity in systems with fractal isochrons},
  journal = {Physica D: Nonlinear Phenomena},
  year = {2015},
  volume = {308},
  pages = {40--51},
  doi = {10.1016/j.physd.2015.06.004}
}

@article{tokuda_inferring_2007,
  author = {Tokuda, I. T. and Jain, S. and Kiss, I. Z. and Hudson, J. L.},
  title = {Inferring Phase Equations from Multivariate Time Series},
  journal = {Physical Review Letters},
  year = {2007},
  volume = {99},
  eid = {064101},
  doi = {10.1103/PhysRevLett.99.064101}
}

@article{kralemann_phase_2008,
  author = {Kralemann, B. and Cimponeriu, L. and Rosenblum, M. and Pikovsky, A. and Mrowka, R.},
  title = {Phase dynamics of coupled oscillators reconstructed from data},
  journal = {Physical Review E},
  year = {2008},
  volume = {77},
  eid = {066205},
  doi = {10.1103/PhysRevE.77.066205}
}

@article{onojima_dynamical_2018,
  author = {Onojima, T. and Goto, T. and Mizuhara, H. and Aoyagi, T.},
  title = {A dynamical systems approach for estimating phase interactions between rhythms of different frequencies from experimental data},
  journal = {PLOS Computational Biology},
  year = {2018},
  volume = {14},
  eid = {e1005928},
  doi = {10.1371/journal.pcbi.1005928}
}

@article{ota_interaction_2020,
  author = {Ota, K. and Aihara, I. and Aoyagi, T.},
  title = {Interaction mechanisms quantified from dynamical features of frog choruses},
  journal = {Royal Society Open Science},
  year = {2020},
  volume = {7},
  eid = {191693},
  doi = {10.1098/rsos.191693}
}

@article{kralemann_vivo_2013,
  author = {Kralemann, B. and Fr{\"u}hwirth, M. and Pikovsky, A. and Rosenblum, M. and Kenner, T. and Schaefer, J. and Moser, M.},
  title = {In vivo cardiac phase response curve elucidates human respiratory heart rate variability},
  journal = {Nature Communications},
  year = {2013},
  volume = {4},
  eid = {2418},
  doi = {10.1038/ncomms3418}
}

@article{arai_interlimb_2024,
  author = {Arai, T. and Ota, K. and Funato, T. and Tsuchiya, K. and Aoyagi, T. and Aoi, S.},
  title = {Interlimb coordination is not strictly controlled during walking},
  journal = {Communications Biology},
  year = {2024},
  volume = {7},
  eid = {1152},
  doi = {10.1038/s42003-024-06843-w}
}

@article{furukawa_bayesian_2025,
  author = {Furukawa, H. and Arai, T. and Funato, T. and Aoi, S. and Aoyagi, T.},
  title = {{Bayesian} estimation of trunk-leg coordination during walking using phase oscillator models},
  journal = {Neuroscience Research},
  year = {2025},
  volume = {215},
  pages = {47--55},
  doi = {10.1016/j.neures.2024.10.002}
}

@article{funato_evaluation_2016,
  author = {Funato, T. and Yamamoto, Y. and Aoi, S. and Imai, T. and Aoyagi, T. and Tomita, N. and Tsuchiya, K.},
  title = {Evaluation of the Phase-Dependent Rhythm Control of Human Walking Using Phase Response Curves},
  journal = {PLOS Computational Biology},
  year = {2016},
  volume = {12},
  eid = {e1004950},
  doi = {10.1371/journal.pcbi.1004950}
}

@article{arai_extracting_2022,
  author = {Arai, T. and Kawamura, Y. and Aoyagi, T.},
  title = {Extracting Phase Coupling Functions between Collectively Oscillating Networks from Time-Series Data},
  journal = {Journal of the Physical Society of Japan},
  year = {2022},
  volume = {91},
  eid = {124001},
  doi = {10.7566/JPSJ.91.124001}
}

@article{arai_setting_2025,
  author = {Arai, T. and Kawamura, Y. and Aoyagi, T.},
  title = {Setting of the {Poincar{\'e}} section for accurately calculating the phase of rhythmic spatiotemporal dynamics},
  journal = {Physical Review E},
  year = {2025},
  volume = {111},
  eid = {014205},
  doi = {10.1103/PhysRevE.111.014205}
}

@misc{arai_data_driven_2026,
  author = {Arai, T. and Aoyagi, T. and Kawamura, Y.},
  title = {Data-driven reconstruction of spatiotemporal phase dynamics for traveling and oscillating patterns via {Bayesian} inference},
  year = {2026},
  publisher = {arXiv},
  howpublished = {2604.23727},
  doi = {10.48550/arXiv.2604.23727}
}

@article{shirasaka_phase-amplitude_2017,
  author = {Shirasaka, S. and Kurebayashi, W. and Nakao, H.},
  title = {Phase-amplitude reduction of transient dynamics far from attractors for limit-cycling systems},
  journal = {Chaos},
  year = {2017},
  volume = {27},
  eid = {023119},
  doi = {10.1063/1.4977195}
}

@article{mauroy_global_2018,
  author = {Mauroy, A. and Mezi{\'c}, I.},
  title = {Global computation of phase-amplitude reduction for limit-cycle dynamics},
  journal = {Chaos},
  year = {2018},
  volume = {28},
  eid = {073108},
  doi = {10.1063/1.5030175}
}

@misc{matsuki_isochron_free_2026,
  author = {Matsuki, A. and Kobayashi, R. and Kori, H.},
  title = {An Isochron-Free Framework for Phase Reduction and Coupling Inference},
  year = {2026},
  publisher = {arXiv},
  howpublished = {2606.25892},
  doi = {10.48550/arXiv.2606.25892}
}

@article{yawata_phase_2024,
  author = {Yawata, K. and Fukami, K. and Taira, K. and Nakao, H.},
  title = {Phase autoencoder for limit-cycle oscillators},
  journal = {Chaos},
  year = {2024},
  volume = {34},
  eid = {063111},
  doi = {10.1063/5.0205718}
}

@article{rosenblum_detecting_2001,
  author = {Rosenblum, M. G. and Pikovsky, A. S.},
  title = {Detecting direction of coupling in interacting oscillators},
  journal = {Physical Review E},
  year = {2001},
  volume = {64},
  eid = {045202(R)},
  doi = {10.1103/PhysRevE.64.045202}
}

@article{miyazaki_determination_2006,
  author = {Miyazaki, J. and Kinoshita, S.},
  title = {Determination of a Coupling Function in Multicoupled Oscillators},
  journal = {Physical Review Letters},
  year = {2006},
  volume = {96},
  eid = {194101},
  doi = {10.1103/PhysRevLett.96.194101}
}

@article{acebron_kuramoto_2005,
  author = {Acebr{\'o}n, J. A. and Bonilla, L. L. and P{\'e}rez Vicente, C. J. and Ritort, F. and Spigler, R.},
  title = {The {Kuramoto} model: A simple paradigm for synchronization phenomena},
  journal = {Reviews of Modern Physics},
  year = {2005},
  volume = {77},
  number = {1},
  pages = {137--185},
  doi = {10.1103/RevModPhys.77.137}
}

@article{cestnik_inferring_2018,
  author = {Cestnik, R. and Rosenblum, M.},
  title = {Inferring the phase response curve from observation of a continuously perturbed oscillator},
  journal = {Scientific Reports},
  year = {2018},
  volume = {8},
  eid = {13606},
  doi = {10.1038/s41598-018-32069-y}
}

@article{wilson_isostable_2016,
  author = {Wilson, D. and Moehlis, J.},
  title = {Isostable reduction of periodic orbits},
  journal = {Physical Review E},
  year = {2016},
  volume = {94},
  eid = {052213},
  doi = {10.1103/PhysRevE.94.052213}
}

@article{stankovski_inference_2012,
  author = {Stankovski, T. and Duggento, A. and McClintock, P. V. E. and Stefanovska, A.},
  title = {Inference of Time-Evolving Coupled Dynamical Systems in the Presence of Noise},
  journal = {Physical Review Letters},
  year = {2012},
  volume = {109},
  eid = {024101},
  doi = {10.1103/PhysRevLett.109.024101}
}

@article{duggento_dynamical_2012,
  author = {Duggento, A. and Stankovski, T. and McClintock, P. V. E. and Stefanovska, A.},
  title = {Dynamical {Bayesian} inference of time-evolving interactions: From a pair of coupled oscillators to networks of oscillators},
  journal = {Physical Review E},
  year = {2012},
  volume = {86},
  eid = {061126},
  doi = {10.1103/PhysRevE.86.061126}
}

@misc{ota_direct_2014,
  author = {Ota, K. and Aoyagi, T.},
  title = {Direct extraction of phase dynamics from fluctuating rhythmic data based on a {Bayesian} approach},
  year = {2014},
  publisher = {arXiv},
  howpublished = {1405.4126},
  doi = {10.48550/arXiv.1405.4126}
}

@article{luchinsky_inferential_2008,
  author = {Luchinsky, D. G. and Smelyanskiy, V. N. and Duggento, A. and McClintock, P. V. E.},
  title = {Inferential framework for nonstationary dynamics. {I}. {Theory}},
  journal = {Physical Review E},
  year = {2008},
  volume = {77},
  eid = {061105},
  doi = {10.1103/PhysRevE.77.061105}
}

@book{bishop_pattern_2006,
  author = {Bishop, C. M.},
  title = {Pattern Recognition and Machine Learning},
  publisher = {Springer},
  address = {New York},
  year = {2006},
  isbn = {978-0-387-31073-2}
}

@book{redner_guide_2001,
  author = {Redner, S.},
  title = {A Guide to First-Passage Processes},
  publisher = {Cambridge University Press},
  address = {Cambridge},
  year = {2001},
  doi = {10.1017/CBO9780511606014}
}

@article{duggento_inferential_2008,
  author = {Duggento, A. and Luchinsky, D. G. and Smelyanskiy, V. N. and Khovanov, I. and McClintock, P. V. E.},
  title = {Inferential framework for nonstationary dynamics. {II}. {Application} to a model of physiological signaling},
  journal = {Physical Review E},
  year = {2008},
  volume = {77},
  eid = {061106},
  doi = {10.1103/PhysRevE.77.061106}
}

@book{kloeden_numerical_1992,
  author = {Kloeden, P. E. and Platen, E.},
  title = {Numerical Solution of Stochastic Differential Equations},
  publisher = {Springer},
  address = {Berlin, Heidelberg},
  year = {1992},
  doi = {10.1007/978-3-662-12616-5}
}

@article{mackay_bayesian_1992,
  author = {MacKay, D. J. C.},
  title = {{Bayesian} interpolation},
  journal = {Neural Computation},
  year = {1992},
  volume = {4},
  number = {3},
  pages = {415--447},
  doi = {10.1162/neco.1992.4.3.415}
}

@article{koopman_hamiltonian_1931,
  author  = {Koopman, B. O.},
  title   = {Hamiltonian systems and transformation in {Hilbert} space},
  journal = {Proceedings of the National Academy of Sciences of the United States of America},
  year    = {1931},
  volume  = {17},
  number  = {5},
  pages   = {315--318},
  doi     = {10.1073/pnas.17.5.315}
}

@article{mezic_spectral_2005,
  author  = {Mezi{\'c}, I.},
  title   = {Spectral properties of dynamical systems, model reduction and decompositions},
  journal = {Nonlinear Dynamics},
  year    = {2005},
  volume  = {41},
  pages   = {309--325},
  doi     = {10.1007/s11071-005-2824-x}
}

@article{rowley_spectral_nonlinear_flows_2009,
  author = {Rowley, C. W. and Mezi{\'c}, I. and Bagheri, S. and Schlatter, P. and Henningson, D. S.},
  title = {Spectral analysis of nonlinear flows},
  journal = {Journal of Fluid Mechanics},
  year = {2009},
  volume = {641},
  pages = {115--127},
  doi = {10.1017/S0022112009992059}
}

@article{schmid_dynamic_mode_decomposition_2010,
  author = {Schmid, P. J.},
  title = {Dynamic mode decomposition of numerical and experimental data},
  journal = {Journal of Fluid Mechanics},
  year = {2010},
  volume = {656},
  pages = {5--28},
  doi = {10.1017/S0022112010001217}
}

@article{ichinaga_pydmd_2024,
  author = {Ichinaga, S. M. and Andreuzzi, F. and Demo, N. and Tezzele, M. and Lapo, K. and Rozza, G. and Brunton, S. L. and Kutz, J. N.},
  title = {{PyDMD}: A {Python} package for robust dynamic mode decomposition},
  journal = {Journal of Machine Learning Research},
  year = {2024},
  volume = {25},
  number = {417},
  pages = {1--9},
  url = {https://jmlr.org/papers/v25/24-0739.html}
}

@article{williams_data_driven_2015,
  author  = {Williams, M. O. and Kevrekidis, I. G. and Rowley, C. W.},
  title   = {A data-driven approximation of the {Koopman} operator: Extending dynamic mode decomposition},
  journal = {Journal of Nonlinear Science},
  year    = {2015},
  volume  = {25},
  pages   = {1307--1346},
  doi     = {10.1007/s00332-015-9258-5}
}

@article{brunton_modern_2022,
  author  = {Brunton, S. L. and Budi{\v s}i{\'c}, M. and Kaiser, E. and Kutz, J. N.},
  title   = {Modern {Koopman} theory for dynamical systems},
  journal = {SIAM Review},
  year    = {2022},
  volume  = {64},
  number  = {2},
  pages   = {229--340},
  doi     = {10.1137/21M1401243}
}

@article{giannakis_nonlinear_2012,
  author  = {Giannakis, D. and Majda, A. J.},
  title   = {Nonlinear {Laplacian} spectral analysis for time series with intermittency and low-frequency variability},
  journal = {Proceedings of the National Academy of Sciences of the United States of America},
  year    = {2012},
  volume  = {109},
  number  = {7},
  pages   = {2222--2227},
  doi     = {10.1073/pnas.1118984109}
}

@article{giannakis_data_driven_2019,
  author  = {Giannakis, D.},
  title   = {Data-driven spectral decomposition and forecasting of ergodic dynamical systems},
  journal = {Applied and Computational Harmonic Analysis},
  year    = {2019},
  volume  = {47},
  pages   = {338--396},
  doi     = {10.1016/j.acha.2017.09.001}
}

@article{das_reproducing_2021,
  author  = {Das, S. and Giannakis, D. and Slawinska, J.},
  title   = {Reproducing kernel {Hilbert} space compactification of unitary evolution groups},
  journal = {Applied and Computational Harmonic Analysis},
  year    = {2021},
  volume  = {54},
  pages   = {75--136},
  doi     = {10.1016/j.acha.2021.02.004}
}

@article{froyland_spectral_2021,
  author  = {Froyland, G. and Giannakis, D. and Lintner, B. R. and Pike, M. and Slawinska, J.},
  title   = {Spectral analysis of climate dynamics with operator-theoretic approaches},
  journal = {Nature Communications},
  year    = {2021},
  volume  = {12},
  eid   = {6570},
  doi     = {10.1038/s41467-021-26357-x}
}

@article{alexander_kernel_analog_forecasting_2017,
  author = {Alexander, R. and Zhao, Z. and Sz{\'e}kely, E. and Giannakis, D.},
  title = {Kernel analog forecasting of tropical intraseasonal oscillations},
  journal = {Journal of the Atmospheric Sciences},
  year = {2017},
  volume = {74},
  number = {4},
  pages = {1321--1342},
  doi = {10.1175/JAS-D-16-0147.1}
}

@article{lintner_identification_2023,
  author  = {Lintner, B. R. and Giannakis, D. and Pike, M. and Slawinska, J.},
  title   = {Identification of the {Madden--Julian Oscillation} with data-driven {Koopman} spectral analysis},
  journal = {Geophysical Research Letters},
  year    = {2023},
  volume  = {50},
  number  = {10},
  eid   = {e2023GL102743},
  doi     = {10.1029/2023GL102743}
}

@article{valva_qbo_2025,
  author  = {Valva, C. and Gerber, E. P.},
  title   = {The {QBO}, the annual cycle, and their interactions: Isolating periodic modes with {Koopman} analysis},
  journal = {Journal of Climate},
  year    = {2025},
  volume  = {38},
  pages   = {3437--3452},
  doi     = {10.1175/JCLI-D-24-0411.1}
}

@book{brunton_data_driven_2022,
  author = {Brunton, S. L. and Kutz, J. N.},
  title = {Data-Driven Science and Engineering: Machine Learning, Dynamical Systems, and Control},
  year = {2022},
  publisher = {Cambridge University Press},
  doi = {10.1017/9781009089517},
  isbn = {9781009098489}
}

@article{timmermann_el_nino_southern_oscillation_complexity_2018,
  author = {Timmermann, Axel and An, Soon-Il and Kug, Jong-Seong and Jin, Fei-Fei and Cai, Wenju and Capotondi, Antonietta and Cobb, Kim M. and Lengaigne, Matthieu and McPhaden, Michael J. and Stuecker, Malte F. and Stein, Karl and Wittenberg, Andrew T. and Yun, Kyung-Sook and Bayr, Tobias and Chen, Han-Ching and Chikamoto, Yoshimitsu and Dewitte, Boris and Dommenget, Dietmar and Grothe, Pamela and Guilyardi, Eric and Ham, Yoo-Geun and Hayashi, Michiya and Ineson, Sarah and Kang, Daehyun and Kim, Sunyong and Kim, WonMoo and Lee, June-Yi and Li, Tim and Luo, Jing-Jia and McGregor, Shayne and Planton, Yann and Power, Scott and Rashid, Harun and Ren, Hong-Li and Santoso, Agus and Takahashi, Ken and Todd, Alexander and Wang, Guomin and Wang, Guojian and Xie, Ruihuang and Yang, Woo-Hyun and Yeh, Sang-Wook and Yoon, Jinho and Zeller, Elke and Zhang, Xuebin},
  title = {{El Ni{\~n}o--Southern Oscillation} complexity},
  journal = {Nature},
  year = {2018},
  volume = {559},
  number = {7715},
  pages = {535--545},
  doi = {10.1038/s41586-018-0252-6}
}

@article{neelin_enso_theory_1998,
  author = {Neelin, J. D. and Battisti, D. S. and Hirst, A. C. and Jin, F.-F. and Wakata, Y. and Yamagata, T. and Zebiak, S. E.},
  title = {{ENSO} theory},
  journal = {Journal of Geophysical Research: Oceans},
  year = {1998},
  volume = {103},
  number = {C7},
  pages = {14261--14290},
  doi = {10.1029/97JC03424}
}

@article{jin_equatorial_ocean_recharge_paradigm_1997,
  author = {Jin, Fei-Fei},
  title = {An equatorial ocean recharge paradigm for {ENSO}. {Part I}: Conceptual model},
  journal = {Journal of the Atmospheric Sciences},
  year = {1997},
  volume = {54},
  number = {7},
  pages = {811--829},
  doi = {10.1175/1520-0469(1997)054<0811:AEORPF>2.0.CO;2}
}

@article{burgers_simplest_enso_recharge_oscillator_2005,
  author = {Burgers, G. and Jin, F.-F. and van Oldenborgh, G. J.},
  title = {The simplest {ENSO} recharge oscillator},
  journal = {Geophysical Research Letters},
  year = {2005},
  volume = {32},
  number = {13},
  eid = {L13706},
  doi = {10.1029/2005GL022951}
}

@article{vialard_enso_recharge_oscillator_conceptual_model_2025,
  author = {Vialard, J. and Jin, F.-F. and McPhaden, M. J. and Fedorov, A. and Cai, W. and An, S.-I. and Dommenget, D. and Fang, X. and Stuecker, M. F. and Wang, C. and Wittenberg, A. and Zhao, S. and Liu, F. and Kim, S.-K. and Planton, Y. and Geng, T. and Lengaigne, M. and Capotondi, A. and Chen, N. and Geng, L. and Hu, S. and Izumo, T. and Kug, J.-S. and Luo, J.-J. and McGregor, S. and Pagli, B. and Priya, P. and Stevenson, S. and Thual, S.},
  title = {The {El Ni{\~n}o Southern Oscillation} ({ENSO}) recharge oscillator conceptual model: Achievements and future prospects},
  journal = {Reviews of Geophysics},
  year = {2025},
  volume = {63},
  number = {1},
  eid = {e2024RG000843},
  doi = {10.1029/2024RG000843}
}

@article{stein_phase_synchronization_2011,
  author = {Stein, K. and Timmermann, A. and Schneider, N.},
  title = {Phase synchronization of the {El Ni{\~n}o--Southern Oscillation} with the annual cycle},
  journal = {Physical Review Letters},
  year = {2011},
  volume = {107},
  eid = {128501},
  doi = {10.1103/PhysRevLett.107.128501}
}

@article{stein_enso_seasonal_synchronization_theory_2014,
  author = {Stein, K. and Timmermann, A. and Schneider, N. and Jin, F.-F. and Stuecker, M. F.},
  title = {{ENSO} seasonal synchronization theory},
  journal = {Journal of Climate},
  year = {2014},
  volume = {27},
  pages = {5285--5310},
  doi = {10.1175/JCLI-D-13-00525.1}
}

@article{stein_seasonal_synchronization_enso_events_2010,
  author = {Stein, K. and Schneider, N. and Timmermann, A. and Jin, F.-F.},
  title = {Seasonal synchronization of {ENSO} events in a linear stochastic model},
  journal = {Journal of Climate},
  year = {2010},
  volume = {23},
  pages = {5629--5643},
  doi = {10.1175/2010JCLI3292.1}
}

@article{chen_dynamics_enso_phase_locking_2022,
  author = {Chen, H.-C. and Jin, F.-F.},
  title = {Dynamics of {ENSO} phase-locking and its biases in climate models},
  journal = {Geophysical Research Letters},
  year = {2022},
  volume = {49},
  number = {3},
  eid = {e2021GL097603},
  doi = {10.1029/2021GL097603}
}

@article{an_linear_solutions_frequency_amplitude_modulation_2011,
  author = {An, S.-I. and Jin, F.-F.},
  title = {Linear solutions for the frequency and amplitude modulation of {ENSO} by the annual cycle},
  journal = {Tellus A},
  year = {2011},
  volume = {63},
  number = {2},
  pages = {238--243},
  doi = {10.1111/j.1600-0870.2010.00482.x}
}

@article{jin_el_nino_devils_staircase_1994,
  author = {Jin, F.-F. and Neelin, J. D. and Ghil, M.},
  title = {{El Ni{\~n}o} on the devil's staircase: Annual subharmonic steps to chaos},
  journal = {Science},
  year = {1994},
  volume = {264},
  number = {5155},
  pages = {70--72},
  doi = {10.1126/science.264.5155.70}
}

@article{minobe_generation_interannual_interdecadal_climate_oscillations_2004,
  author = {Minobe, S. and Jin, F.-F.},
  title = {Generation of interannual and interdecadal climate oscillations through nonlinear subharmonic resonance in delayed oscillators},
  journal = {Geophysical Research Letters},
  year = {2004},
  volume = {31},
  number = {16},
  eid = {L16206},
  doi = {10.1029/2004GL019776}
}

@article{rasmusson_variations_tropical_sea_surface_temperature_1982,
  author = {Rasmusson, E. M. and Carpenter, T. H.},
  title = {Variations in tropical sea surface temperature and surface wind fields associated with the {Southern Oscillation/El Ni{\~n}o}},
  journal = {Monthly Weather Review},
  year = {1982},
  volume = {110},
  number = {5},
  pages = {354--384},
  doi = {10.1175/1520-0493(1982)110<0354:VITSST>2.0.CO;2}
}

@article{zebiak_model_el_nino_southern_oscillation_1987,
  author = {Zebiak, S. E. and Cane, M. A.},
  title = {A model {El Ni{\~n}o--Southern Oscillation}},
  journal = {Monthly Weather Review},
  year = {1987},
  volume = {115},
  number = {10},
  pages = {2262--2278},
  doi = {10.1175/1520-0493(1987)115<2262:AMENO>2.0.CO;2}
}

@article{bjerknes_atmospheric_teleconnections_1969,
  author = {Bjerknes, J.},
  title = {Atmospheric teleconnections from the equatorial {Pacific}},
  journal = {Monthly Weather Review},
  year = {1969},
  volume = {97},
  number = {3},
  pages = {163--172},
  doi = {10.1175/1520-0493(1969)097<0163:ATFTEP>2.3.CO;2}
}

@article{horel_planetary_scale_atmospheric_phenomena_1981,
  author = {Horel, J. D. and Wallace, J. M.},
  title = {Planetary-scale atmospheric phenomena associated with the {Southern Oscillation}},
  journal = {Monthly Weather Review},
  year = {1981},
  volume = {109},
  number = {4},
  pages = {813--829},
  doi = {10.1175/1520-0493(1981)109<0813:PSAPAW>2.0.CO;2}
}

@article{ropelewski_global_regional_precipitation_patterns_1987,
  author = {Ropelewski, C. F. and Halpert, M. S.},
  title = {Global and regional scale precipitation patterns associated with the {El Ni{\~n}o/Southern Oscillation}},
  journal = {Monthly Weather Review},
  year = {1987},
  volume = {115},
  number = {8},
  pages = {1606--1626},
  doi = {10.1175/1520-0493(1987)115<1606:GARSPP>2.0.CO;2}
}

@article{torrence_annual_cycle_persistence_1998,
  author = {Torrence, C. and Webster, P. J.},
  title = {The annual cycle of persistence in the {El Ni{\~n}o/Southern Oscillation}},
  journal = {Quarterly Journal of the Royal Meteorological Society},
  year = {1998},
  volume = {124},
  number = {550},
  pages = {1985--2004},
  doi = {10.1002/qj.49712455010}
}

@article{levine_annual_cycle_enso_growth_rate_2015,
  author = {Levine, A. F. Z. and McPhaden, M. J.},
  title = {The annual cycle in {ENSO} growth rate as a cause of the spring predictability barrier},
  journal = {Geophysical Research Letters},
  year = {2015},
  volume = {42},
  number = {12},
  pages = {5034--5041},
  doi = {10.1002/2015GL064309}
}

@article{penland_optimal_growth_1995,
  author = {Penland, C. and Sardeshmukh, P. D.},
  title = {The optimal growth of tropical sea surface temperature anomalies},
  journal = {Journal of Climate},
  year = {1995},
  volume = {8},
  pages = {1999--2024},
  doi = {10.1175/1520-0442(1995)008<1999:TOGOTS>2.0.CO;2}
}

@article{thompson_linear_stochastic_dynamical_model_enso_2001,
  author = {Thompson, C. J. and Battisti, D. S.},
  title = {A linear stochastic dynamical model of {ENSO}. Part {II}: Analysis},
  journal = {Journal of Climate},
  year = {2001},
  volume = {14},
  pages = {445--466},
  doi = {10.1175/1520-0442(2001)014<0445:ALSDMO>2.0.CO;2}
}

@article{kido_understanding_interplay_enso_2023,
  author = {Kido, S. and Richter, I. and Tozuka, T. and Chang, P.},
  title = {Understanding the interplay between {ENSO} and related tropical {SST} variability using linear inverse models},
  journal = {Climate Dynamics},
  year = {2023},
  volume = {61},
  pages = {1029--1048},
  doi = {10.1007/s00382-022-06484-x}
}

@article{lien_cyclostationary_linear_inverse_models_2025,
  author = {Lien, J. and Kuo, Y.-N. and Ando, H. and Kido, S.},
  title = {On cyclostationary linear inverse models: a mathematical insight and implication},
  journal = {Frontiers in Complex Systems},
  year = {2025},
  volume = {3},
  eid = {1563687},
  doi = {10.3389/fcpxs.2025.1563687}
}

@misc{yasuda_stochastic_thermodynamic_constraint_2026,
  author = {Yasuda, Y. and Kohyama, T.},
  title = {A stochastic-thermodynamic constraint on the seasonal phase locking of the {El Ni{\~n}o-Southern Oscillation}},
  year = {2026},
  publisher = {arXiv},
  howpublished = {2606.19778},
  doi = {10.48550/arXiv.2606.19778}
}

@article{battisti_dynamics_thermodynamics_warming_event_1988,
  author = {Battisti, D. S.},
  title = {Dynamics and thermodynamics of a warming event in a coupled tropical atmosphere--ocean model},
  journal = {Journal of the Atmospheric Sciences},
  year = {1988},
  volume = {45},
  number = {20},
  pages = {2889--2919},
  doi = {10.1175/1520-0469(1988)045<2889:DATOAW>2.0.CO;2}
}

@article{wang_review_enso_theories_2018,
  author = {Wang, C.},
  title = {A review of {ENSO} theories},
  journal = {National Science Review},
  year = {2018},
  volume = {5},
  number = {6},
  pages = {813--825},
  doi = {10.1093/nsr/nwy104}
}

\end{document}